\documentclass{aastex701}
\usepackage[utf8]{inputenc}
\usepackage{textgreek}
\DeclareUnicodeCharacter{2061}{}

\begin{document}

\title{Galaxy evolution in compact groups – III. Structural analysis of galaxies and dynamical state of non-isolated compact groups}

\correspondingauthor{Gissel P. Montaguth}
\email{gissel.pmontaguth@usp.br}

\author[orcid=00009-0003-1364-3590,gname=Gissel, sname=P. Montaguth]{Gissel P. Montaguth}
\affiliation{Instituto de Astronomia, Geof\'isica e Ci\^encias Atmosf\'ericas da Universidade de S\~ao Paulo, Cidade Universit\'ria, CEP:05508-990, S\~ao Paulo, SP, Brazil}
\email{gissel.pmontaguth@usp.br}

\author[orcid=0000-0002-6656-5333,gname='Ana Laura', sname='O'Mill']{Ana Laura O'Mill}
\affiliation{CONICET, Instituto de Astronomía Teorica y Experimental (IATE), Laprida 854, Córdoba X5000BGR, Argentina}
\affiliation{Observatorio Astronómico  de Córdoba  (OAC), Universidad Nacional de Córdoba (UNC), Laprida 854, Córdoba X5000BGR, Argentina}
\email{anaomill@unc.edu.ar}

\author[orcid=0000-0002-5267-9065,gname=Claudia Laura, sname=Mendes de Oliveira]{Claudia Mendes de Oliveira}
\affiliation{Instituto de Astronomia, Geof\'isica e Ci\^encias Atmosf\'ericas da Universidade de S\~ao Paulo, Cidade Universit\'ria, CEP:05508-990, S\~ao Paulo, SP, Brazil}
\email{claudia.oliveira@iag.usp.br}

\author[orcid=0009-0006-0373-8168,gname=Ciria, sname=Lima-Dias]{Ciria Lima-Dias}
\affiliation{Departamento de Astronomía, Universidad de La Serena, Avda. Ra\'ul Bitr\'an 1305, La Serena, Chile}
\email{clima@userena.cl}

\author[orcid=0000-0002-7005-8983,gname=Sergio,sname=Torres-Flores]{Sergio Torres-Flores}
\affiliation{Departamento de Astronomía, Universidad de La Serena, Avda. Ra\'ul Bitr\'an 1305, La Serena, Chile}
\email{sptorres@userena.cl}

\author[orcid=0000-0003-2325-9616,gname=Antonela,sname=Monachesi]{Antonela Monachesi}
\affiliation{Departamento de Astronomía, Universidad de La Serena, Avda. Ra\'ul Bitr\'an 1305, La Serena, Chile}
\email{amonachesi@userena.cl}

\author{D. E. Olave-Rojas}
\affiliation{Departamento de Tecnologías Industriales, Facultad de Ingeniería, Universidad de Talca, Los Niches km 1, Curicó, Chile}
\email{daniela.olave@utalca.cl}

\author{Diego Pallero}
\affiliation{Departamento de F\'isica, Universidad T\'ecnica Federico Santa Mar\'ia, Avenida España 1600, Valpara\'iso, Chile}
\affiliation{Millennium Nucleus for Galaxies (MINGAL)}
\email{palleroastargod@gmail.com}

\author{Pedro K. Humire}
\affiliation{Instituto de Astronomia, Geof\'isica e Ci\^encias Atmosf\'ericas da Universidade de S\~ao Paulo, Cidade Universit\'ria, CEP:05508-990, S\~ao Paulo, SP, Brazil}
\email{pedrohumirer@gmail.com}

\author{Ricardo Demarco}
\affiliation{Institute of Astrophysics, Facultad de Ciencias Exactas, Universidad Andr\'es Bello, Sede Concepci\'on, Talcahuano, Chile}
\email{demarco.rj@gmail.com}

\author{Eduardo Telles}
\affiliation{Observat\'orio Nacional, Rua General Jos\'e Cristino, 77,  S\~ao Crist\'ov\~ao, 20921-400 Rio de Janeiro, RJ, Brazil}
\email{etelles@on.br}

\author{Paulo A. A. Lopes}
\affiliation{Observat\'orio do Valongo, Universidade Federal do Rio de Janeiro, Ladeira Pedro Ant\^{o}nio 43, Rio de Janeiro, RJ, 20080-090, Brazil}
\email{plopes@ov.ufrj.br}

\author{Swayamtrupta Panda}
\affiliation{International Gemini Observatory/NSF NOIRLab, Casilla 603, La Serena, Chile}
\email{swayamtrupta.panda@noirlab.edu}

\author{Rodrigo F. Haack}
\affiliation{Instituto de Astrofísica de La Plata, UNLP-CONICET,  Paseo del Bosque s/n, La Plata, B1900FWA, Argentina}
\affiliation{Facultad de Ciencias Astronómicas y Geofísicas, Universidad Nacional de La Plata, Paseo del Bosque s/n, La Plata, B1900FWA, Argentina}
\email{rodrihaack@gmail.com}

\author{Amanda R. Lopes}
\affiliation{Instituto de Astrofísica de La Plata, UNLP-CONICET,  Paseo del Bosque s/n, La Plata, B1900FWA, Argentina}
\email{amandalopes1920@gmail.com}

\author{Alvaro Alvarez-Candal}
\affiliation{Instituto de Astrof\'isica de Andaluc\'ia, CSIC, Apt 3004, E18080 Granada, Spain}
\email{varobes@gmail.com}

\author{Analia V. Smith Castelli}
\affiliation{Instituto de Astrofísica de La Plata, UNLP-CONICET,  Paseo del Bosque s/n, La Plata, B1900FWA, Argentina}
\affiliation{Facultad de Ciencias Astronómicas y Geofísicas, Universidad Nacional de La Plata, Paseo del Bosque s/n, La Plata, B1900FWA, Argentina}
\email{a.smith.castelli@gmail.com}

\author{Antonio Kanaan}
\affiliation{Departamento de F\'isica, Universidade Federal de Santa Catarina, Florian\'opolis, SC, 88040-900, Brazil}
\email{ankanaan@gmail.com}

\author{Tiago Ribeiro}
\affiliation{NOAO, 950 North Cherry Ave. Tucson, AZ 85719, United States}
\email{tiago.astro@gmail.com}

\author{William Schoenell}
\affiliation{GMTO Corporation, N. Halstead Street 465, Suite 250, Pasadena, CA 91107, United States}
\email{wschoenell@gmail.com}

\begin{abstract}

Compact Groups (CGs) of galaxies are dense systems where projected separations are comparable to their optical diameters. A subset — non-isolated CGs — are embedded within major structures. Using multi-band S-PLUS data, we analyse galaxies in 122 non-isolated CGs within more massive systems such as larger groups and clusters. We compare them to galaxies in the host structures, hereafter surrounding group galaxies. Structural parameters were obtained with MorphoPLUS, a pipeline for multi-wavelength Sérsic profile fitting. Dividing galaxies into early (ETG), transition, or late types (LTG), we find: (1) Non-isolated CGs host higher quenched fractions and more ETGs, especially for stellar masses $\log(M/M_\odot) > 10.2$, than surrounding groups. (2) Sérsic indices increase with wavelength for all morphological types in both environments, whereas effective radii show a stronger morphology-dependent behaviour — ETGs become more compact towards redder bands, while LTGs exhibit flatter 
$Re(\lambda)$ trends. Environmental differences remain weak, with only a modest enhancement of the gradients for ETGs in non-isolated CGs. (3) Transition galaxies in CGs show a concentrated $R_e$–$n$ distribution and faint-end bimodality, consistent with ongoing morphological transformation absent in surrounding groups. (4) Phase-space analysis indicates that some CGs in clusters are projection artefacts, while others are genuine dense systems at various infall stages, from recent arrivals to ancient remnants. These results show that galaxies in non-isolated CGs follow distinct evolutionary paths compared to their surrounding groups galaxies, suggesting that the compact configuration plays a unique role beyond the influence of the larger-scale environment.

\end{abstract}


\section{Introduction}

Compact groups (CGs) of galaxies are key systems for studying gravitational interactions and their impact on galaxy evolution. These systems exhibit high densities, similar to the cores of galaxy clusters, but with much lower velocity dispersions, leading to more frequent and intense close encounters (\citealt{1982Hickson}). 
CGs exhibit distinctive properties compared to less dense environments, with member galaxies that are preferentially more compact, redder, and with early-type morphologies (\citealt{2004lee}, \citealt{2008deng}, \citealt{2012Coenda}, \citealt{Poliakov_2021}, \citealt{2023Montaguth}).
Additionally, CGs often contain diffuse intra-group light from disrupted galaxies (\citealt{2005DaRocha}, \citealt{Poliakov_2021}) 
and display a unique distribution in the mid-infrared 3.6-8.0 micronum colour space, with a clear bimodal distribution and a gap not seen in any other environment, a feature known as the "canyon", which suggests a rapid transition from star-forming to quiescent state for galaxies in CGs 
(\citealt{2007AJohnson}, \citealt{2008Gallagher}, \citealt{2010Walker}, \citeyear{2013Walker}). These characteristics highlight CGs as crucial environments for studying galaxy transformation processes and for understanding the preprocessing of galaxies within galaxy groups.      
                                                            
In the early days of CG, \cite{rose1977survey} found that almost every CG had at least one galaxy closer to it than expected for randomly distributed field galaxies, indicating that most CGs are small subsystems within larger structures. In addition, \cite{1998Ribeiro} analysed 17 CGs identified by \citealt{1982Hickson} and found significant differences in the surface density of these systems and their surroundings, suggesting that CGs may exist in different dynamical stages. In their study, the CGs were classified into three categories: real compact systems, core-halo systems, and loose groups. This classification reinforces the idea that CGs are not a homogeneous family of objects but rather represent different evolutionary stages. In fact, CGs have been detected embedded in a variety of environments within the galaxies' large-scale structure, from voids to the most massive clusters (\citealt{2010Diaz}, \citealt{2021Zheng}, \citealt{2022Taverna}, \citealt{2023Taverna}).

This raises a key question on how the local environment and large-scale structure influence the evolution of CGs. Although the influence of environment on galaxy evolution has been widely studied, the connection between local and large-scale structures and its overall impact on galaxies remains largely unexplored. Recent studies have approached this issue from different perspectives, analysing the properties of galaxies in voids (\citealt{2023Rodriguez}), and groups in different large-scale structure environments (\citealt{2024Torres-Rios}). These studies consistently find that the properties of galaxies in groups are influenced by their position within the cosmic web.

In the specific case of CGs, \cite{2023Taverna} found that their physical properties vary depending on their environments: CGs located in dense regions, such as nodes, exhibit the highest velocity dispersions, the most luminous first-ranked galaxies, and the shortest crossing times, whereas non-embedded, isolated CGs show the opposite trends. Moreover, when comparing galaxies within and around CGs across different environments, CGs consistently contain higher fractions of red and early-type members than their surrounding groups (\citealt{2023Taverna}). Consistent with these findings, \cite{2024Montaguth} and \cite{2024Zandivarez} also found that CGs embedded in major structures exhibit distinct morphological characteristics and star formation rates compared to isolated CGs, suggesting that the environment has been key to shaping the evolution of its galaxy members and witnessing different evolutionary stages.


The dynamics of CGs have been widely studied both in simulations and observations. The works of \citet{2020Diaz} and \citet{2022Taverna} find that between 35\% and 65\% of the CGs may be dense 3D systems, either isolated or embedded within larger structures.  \citet{2021Zheng}, using observations and monte-carlo simulations, find that for embedded CGs, the relationship between their velocity dispersions and those of their host groups indicates that some CGs have distinct dynamics and thus may be newly accreted substructures, whereas others follow the dynamics of the groups they belong to.

In this context, a key question is whether the evolution of galaxies in non-isolated CGs is influenced by the structure and dynamics of their larger host groups. To address this, we compare the properties of galaxies in non-isolated CGs with those of galaxies in their surrounding host groups. 
To achieve this,  we will study a sample of host groups and clusters that contain 122 non-isolated CGs studied in \cite{2024Montaguth}.
We make use of the published morphological parameters of the galaxies in these 122 non-isolated CGs, estimated by \cite{2023Montaguth} using images from the Southern Photometric Local Universe Survey (S-PLUS) project (\citealt{mendes2019southern}), which provides photometric information across 12 bands, enabling a precise and detailed multi-wavelength analysis. In this paper, we further analyse the morphologies of galaxies of the host groups and clusters that contain these 122 non-isolated CGs using S-PLUS data and a new routine \textsc{MorphoPLUS}, exactly on the same way as done in \cite{2023Montaguth}. Additionally, we will use the GALEX-SDSS-WISE LEGACY catalogue (\citealt{2018GSWLC}) to investigate star-formation rates and stellar masses of the galaxies in all samples.

This paper is structured as follows: in Section \ref{sec:data}, we provide a detailed description of the data that will be used in this study, including the criteria for selecting the CGs and their surrounding structures. Section \ref{sec:methodology} details the methodology we use to derive the structural parameters of the galaxies through a comprehensive multi-wavelength analysis, as well as to obtain the specific star formation rate. Our findings are presented and discussed in Sections \ref{sec:results} and \ref{sec:discussion}, respectively. Finally, in Section \ref{sec:conclusions}, we summarize the key outcomes of our work and draw our conclusions. Throughout the paper, we adopt a flat cosmological model with parameters $H_0 = 70 km$ $s^{-1}$ $Mpc^{-1}$, $\Omega_M = 0.3$, and $\Omega_\lambda = 0.7$ (\citealt{2003Spergel}).

\section{Data}
\label{sec:data}

In this section, we briefly describe the selection of CGs and their environmental characterization. In \cite{2023Montaguth}, we selected 316 CGs in the Stripe 82 region (\citealt{2009Dr7}) using the catalogue by \cite{zheng2020compact} and redshifts from SDSS-DR14 (\citealt{2018Sdss14}), LAMOST (\citealt{2015Lamost}), and GAMA (\citealt{2015GAMA}). The selection criteria combined Hickson’s photometric approach (\citealt{1982Hickson}) with a spectroscopic method, which requires a radial velocity difference of $\Delta v \leq 1000$ km/s from the mean radial velocity of the CG and its member galaxies. The photometric criterion considers the group’s surface brightness, which is determined by the total magnitude of all galaxies averaged over the smallest circle that encloses them. This value must be lower than 26 mag arcsec$^{-2}$ in the r-band. Additionally, all member galaxies must have apparent magnitudes within $14 \leq r \leq 17.7$. The isolation criterion requires $\theta_{N} \geq 3\theta_{G}$, where $\theta_{G}$ is the radius of the group and $\theta_{N}$ is the angular distance to the brightest non-member galaxy closest to the group.

Our analysis is explicitly based on Hickson-like CGs, as the observed properties of CGs depend sensitively on the selection method, with each methodology introducing its own biases \citep{Hernández-Fernández2014,2016Taverna}. The classical isolation criterion proposed by \citet{1982Hickson} was intended to minimise contamination by interlopers in a photometrically selected sample of apparently isolated systems. However, subsequent spectroscopic studies revealed that a significant fraction of Hickson CGs are in fact embedded within looser galaxy associations — that is, they are not truly isolated in the sense of residing alone within a dark matter halo \citep[e.g.][]{1998Ribeiro}.

More recent large catalogues have quantified this picture for Hickson-like samples, showing that roughly half of the CGs are non-isolated or embedded within larger structures such as loose groups or clusters, reinforcing that the isolation criterion is not a reliable indicator of the large-scale environment \citep[e.g.][]{2023Taverna, 2021Zheng}. In agreement with \citet{1996ABarton}, who also argued that the isolation criterion is not a reliable tracer of environment, they showed that when the isolation condition is removed, the resulting samples occupy environments with projected galaxy densities that are statistically similar to those of Hickson CGs, while also revealing a size-dependent selection effect: systems with larger projected radii are more likely to violate the isolation annulus and thus be missed by Hickson-like searches. This introduces two practical biases: (i) nearby compact systems (low redshift) are preferentially excluded because the sky area subtended by the $3\theta_G$ annulus increases, making neighbouring galaxies more likely to appear within it; and (ii) intrinsically larger, embedded CGs are more easily rejected by the isolation condition, this trend was also reported by \citet{2015Sohn, Sohn2016catalogs}.

\citet{Sohn2016catalogs} further showed that groups selected without the isolation criterion differ from Hickson-like CGs in their early-type galaxy fraction. Therefore, to remain consistent with our previous works \citep{2023Montaguth,2024Montaguth} and with the main goal of this study — namely, to determine whether the properties of galaxies in non-isolated Hickson-like CGs differ from those of galaxies in their host environments — our analysis is restricted to Hickson-like, non-isolated CGs.

To further characterize these CGs, we cross-matched them with data release 4 of S-PLUS (DR4; \citealt{2024Herpich}), a survey being performed with a 0.8m telescope at Cerro Tololo Inter-American Observatory in Chile. S-PLUS uses a 12-band filter system, the Javalambre filter system (\citealt{2019Cenarro}), combining broad-band filters (u, g, r, i, z) with narrow-band filters (J0378, J0395, J0410, J0430, J0515, J0660, and J0860). These narrow-band filters are centred on key stellar spectral features, including the Balmer jump/[OII], Ca, H+K, H$_{\delta}$, G-band, Mg b triplet, Hα, and the Ca triplet. We identified 316 CGs (comprising 1083 galaxies) within a redshift range of 0.015–0.197.

In \cite{2024Montaguth}, we analysed the environment in which these CGs are located, following the methodology of \cite{2021Zheng}. For this purpose, we used sample III of the group catalogue by \cite{2007Yang}, which classifies structures using a halo-based method. In this approach, galaxies are assigned to dark matter halos, with the halo radius, $r_{180}$, defined as the radius where the average mass density is 180 times the mean density of the Universe at the corresponding redshift.
Sample III was constructed using all the spectroscopic information available for the SDSS Main Galaxy Sample, and \cite{2007Yang} also incorporated data from external surveys suchas the 2dF Galaxy Redshift Survey \citep{2001Colless}, the Point Source Catalog Redshift Survey \citep{2000Saunders}, and the catalogue of Bright Galaxies \citep{1991DeVaucouleurs}, and by assigning redshifts to SDSS galaxies without spectroscopy using their nearest neighbours.
As noted by \cite{2021Zheng}, this version reaches 100\% completeness of member galaxies until $r\leq17.7$, although it may include a small fraction of contamination from foreground or background galaxies.
To identify the environment of our CGs, we cross-matched the galaxies in the CGs identified by \cite{zheng2020compact} with the galaxies in the groups from \cite{2007Yang}. We found that 278 CGs in our sample (88\%) at least one member galaxy is also listed in the \cite{2007Yang} group catalogue.  From these 278 CGs, in this paper, we focus on 122 non-isolated CGs, embedded within larger structures, meaning that they are subsets of more massive groups or clusters. In \cite{2024Montaguth}, we refer to these as non-isolated CGs. These 122 CGs are distributed across 102 groups identified by \cite{2007Yang}. Additionally, we define a subset of 66 CGs where both \cite{2007Yang} and \cite{zheng2020compact} identify exactly the same member galaxies. We refer to these systems as isolated CGs. This subset will be used as our control sample, and contains a total of 186 galaxies. For further details, we refer to \cite{2024Montaguth}.

As the main goal of this paper is to compare the properties of galaxies in non-isolated CGs with those of galaxies in their host structures, we start by conducting a similar morphological analysis to that done in \cite{2023Montaguth}, using S-PLUS, for the galaxies in the host structures. To accomplish this, we cross-matched the coordinates of 1365 galaxies in the 102 groups with the S-PLUS DR4 catalogue, retrieving all matching sources. From DR4, we obtained magnitudes and errors for the twelve filters, the extinction $E_{B-V}$ from \cite{1998Schlegel}, the position angle, and the elongation of each galaxy. The redshifts were taken from \cite{2007Yang} catalogue. Throughout this paper, we refer to these 102 groups as major structures. Furthermore, we divide the 1365 galaxies within these major structures into two categories: 392 galaxies identified as members of non-isolated CGs and 973 galaxies that are not. The latter group, consisting of galaxies in major structures that are not part of non-isolated CGs, will be referred to as surrounding group galaxies, even if formally a few of the structures could be named clusters, given the large number of objects.

\begin{figure}
    \centering
    \includegraphics[width=0.45\linewidth]{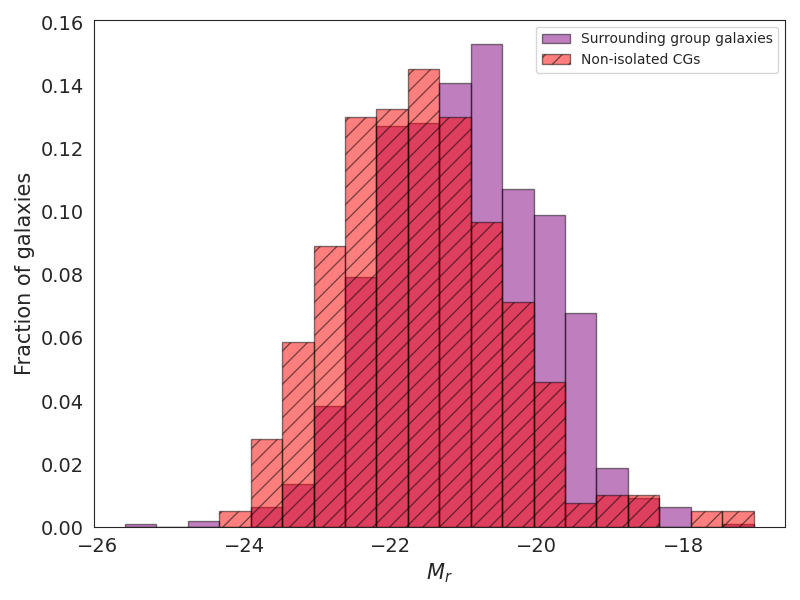}
    \includegraphics[width=0.45\linewidth]{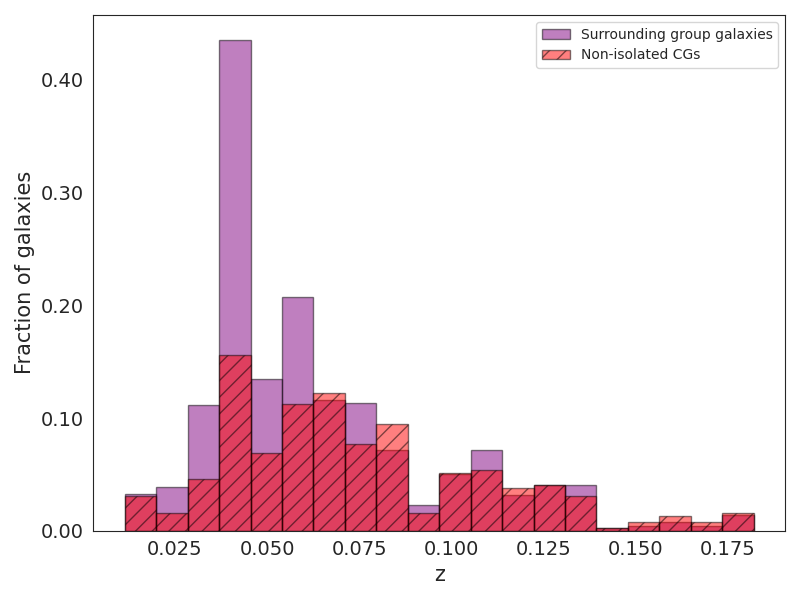}
    \caption{Left: normalised histogram distribution of $M_r$ shows the galaxies in non-isolated CGs represented in red, while those in the surrounding groups (in which the CGs are embedded) are shown in purple. In the latter case, we excluded the galaxies that are part of the CGs. The absolute magnitudes in the $r$-band have been corrected for galactic extinction and include the K-correction. Right: normalised redshift ($z$) distributions for the same samples. }
    \label{fig:Mr}
\end{figure}

In left panel in Figure \ref{fig:Mr}, we present the distribution of absolute magnitudes for galaxies in non-isolated CGs (in red) and for the surrounding group galaxies in the major structures where these non-isolated CGs are embedded (in purple). We estimate the absolute magnitude using the formula $M_r = m_r - 5 \times \log(D_L/10\text{pc}) - K$. Here, $m_r$ is de magnitude in the r band, $D_L$ is the luminosity distance calculated from the redshift, and $K$ is the K-correction. For the K-correction, we employed the publicly available software package by \cite{Blanton2007}, version $V4\_3$, to obtain de-reddened magnitudes at $z = 0$. According to this Figure \ref{fig:Mr} (left panel), galaxies in non-isolated CGs are systematically the brightest members the surrounding groups galaxies. This is largely a selection effect: the compactness  threshold for identifying CGs (brighter than 26 mag arcsec$^{-2}$;\citealt{1982Hickson}) favour intrinsically luminous, high–surface-brightness systems and disfavour low-surface-brightness or extended CGs \citep{2010Diaz,2025Tricottet}. At the same time, this results in the selection of bright galaxies located within larger structures. Splitting the sample into three redshift bins, we confirm that CG galaxies remain brighter than the other galaxies in the major structures, indicating that this is not a redshift-dependent effect. In the results and discussion, we therefore compare the samples within absolute-magnitude bins to mitigate this bias. In the right panel, we present the redshift distribution for both samples, which by construction lie within the same redshift range, $0.015–0.186$.

\section{Methodology}
\label{sec:methodology}

\subsection{\textsc{MorphoPLUS}: morphometric parameters} 
\label{subsec:galfitm}

We develop the \textsc{MorphoPLUS} code\footnote{Available at \url{https://github.com/GMontaguth/MorphoPLUS}}, which automates the extraction of morphometric parameters using S-PLUS data by modeling the surface brightness of galaxies using a Sérsic profile (\citealt{Sersic}). Using this tool, we obtained the Sérsic index ($n$), effective radius ($R_e$), axis ratio ($b/a$), and position angle (PA) for all galaxies in the 12 optical filters. For this, MorphoPLUS uses \textsc{SExtractor} tools (\citealt{bertin1996sextractor}), along with the \textsc{GALFITM} algorithm (\citealt{2011Bamford}, \citealt{2013Haussler},\citealt{2013vika}). \textsc{GALFITM} expands the capabilities of \textsc{GALFIT} 3.02 \citep{2002Peng, 2010Peng} to handle multi-wavelength data, allowing for surface brightness fittings with wavelength-dependent parameters modeled using Chebyshev polynomials. The advantage of this approach is its improved accuracy and robustness in the fit \citep{2013vika}. In Appendix \ref{app:morphoplus}, we provide a more detailed explanation of how the code works.  

In total, we fitted 973 galaxies that are not part of non-isolated CGs in 102 major structures, providing a comprehensive set of morphological parameters for the surrounding group galaxies for further analysis. This was combined with the morphological parameters already obtained for the non-isolated CGs, using the same methodology as here, listed in \cite{2023Montaguth} to complete the study that follows.

\subsubsection{Morphology classification}

\begin{figure}
    \centering
    \includegraphics[scale=0.5]{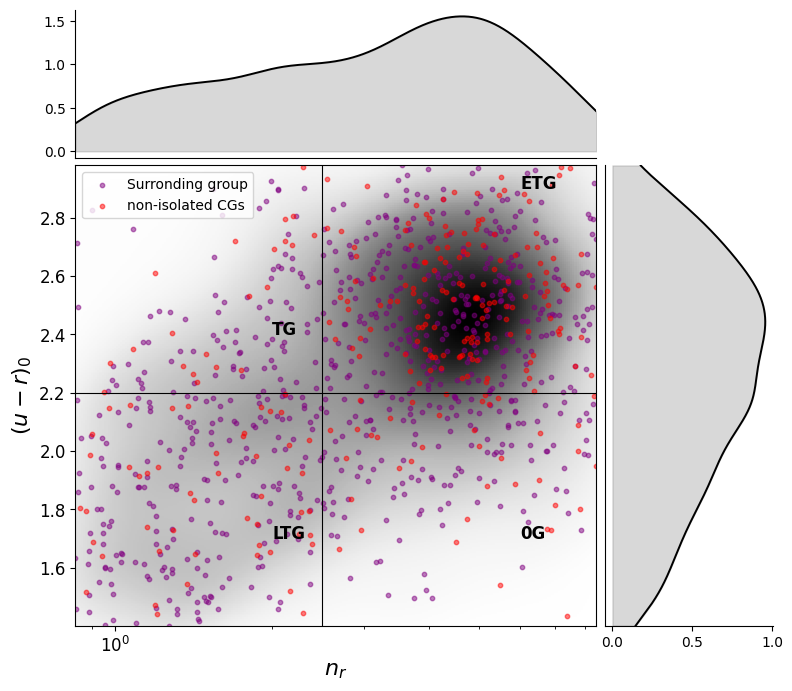}
    \caption{Classification of ETGs, transition galaxies, and LTGs based on rest–frame colour $(u-r)_0$ and the Sérsic index in the r-band ($n_r$). The vertical guide marks ($n_r=2.5$), and the horizontal guide marks $(u-r)_0=2.2$. Purple points show surrounding–group galaxies and red points show galaxies in non–isolated CGs. A greyscale background displays the two–dimensional kernel–density estimate of the joint distribution, and the marginal normalised histograms are shown along the top $(u-r)_0$ axis and the right $log(n_r)$ axis.}

   \label{fig:vika}
\end{figure}

We adopt a colour–structure morphological classification following the rationale of \cite{vika2015megamorph}, which is based on the ($u-r$) colour and the Sérsic index in the $r$-band. Galaxies with $n_r \geq 2.5$ and $(u-r) \geq2.3$ are classified as early-type galaxies (ETGs), while galaxies with $n_r < 2.5$ and $(u-r) < 2.3$ are classified as late-type galaxies (LTGs). Building on this scheme, and as in \cite{2024Montaguth}, we work with the rest–frame colour $(u-r)_0$ (corrected for both k-correction and dust extinction) and determine the operative thresholds directly from the data; the full procedure (1D Gaussian Mixture Models fits, equal–contribution intersections, and robustness checks) is detailed in Appendix~\ref{app:morph_cuts}. We adopt $(u-r)_0=2.2$ and $n_r=2.5$ as our working boundaries.

 In Figure \ref{fig:vika}, 
the purple points show the 973 surrounding group galaxies, while the red points
show the 392 galaxies in non-isolated CGs. 
The horizontal lines in both plots represent the colour $(u-r)_0=2.2$, and the vertical line represents a Sérsic index of $n_r=2.5$. The lower-left quadrant defines our selection of LTGs, the upper-left quadrant contains what we refer to as transition galaxies, the upper-right quadrant contains ETGs, and the lower-right quadrant represents what we label as other galaxies. In Table \ref{tab:morph_mix} we report the percentage of each morphological type; in both environments, ETGs are the largest fraction.

\begin{table}
\centering
\caption{Fractions of ETGs, TGs, LTGs, and Other in each environment (SG vs CG). Classification follows \citet{vika2015megamorph} with $n_r$ and $(u-r)_0$ (K- and dust-corrected).}
\begin{tabular}{lcc}
\hline
 & Surrounding groups (SG; $N=973$) & Non-isolated CGs (CG; $N=392$) \\
\hline
ETGs        & 42\% & 53\% \\
Transition  & 12\%  & 7\%  \\
LTGs        & 29\% & 21\% \\
Other       & 17\% & 19\% \\
\hline
\end{tabular}
\label{tab:morph_mix}
\end{table}

\subsection{Stellar masses and star formation rates}

In order to obtain the stellar masses ($M_*$) and star formation rates (SFR), we used the catalog by \cite{2018GSWLC}, which derived these values through spectral energy distribution (SED) fitting using data from GALEX, SDSS, and WISE. The SED fitting was performed using the CIGALE code (\citealt{2009Noll}). Of the 1365 galaxies in the sample, 1325 are included in the \cite{2018GSWLC} catalog. The remaining 40 galaxies are not included, possibly because in \cite{2018GSWLC} only galaxies with spectroscopic redshifts from SDSS were considered, while \cite{2007Yang} supplemented their data with other catalogs, e.g., 2dFGRS (\citealt{2001Colless}), the PSCz (\citealt{2000Saunders}), and the RC3 (\citealt{1991DeVaucouleurs}). These authors also addressed the well-known issue of fiber collision in SDSS: for those galaxies, the redshift of the nearest neighbor was assigned. We were able to retrieve estimates of specific star formation rates ($sSFR[yr^{-1}]=SFR/M_{*}$) for 1325 galaxies in our samples.

\begin{figure}
    \centering
    \includegraphics[scale=0.55]{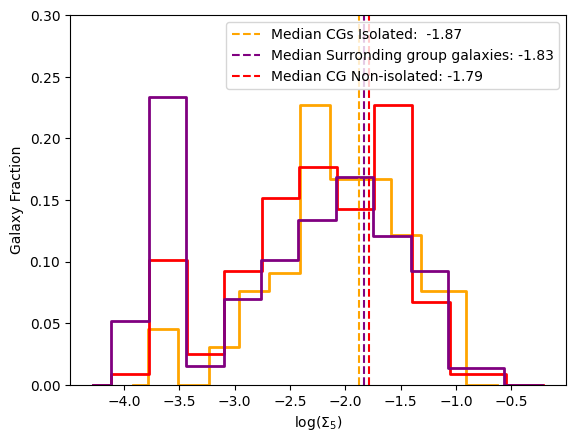}
    \caption{Distributions of projected local density $(\Sigma_5)$. The distributions for the isolated CGs, Non-isolated CGs, and surrounding group galaxies samples are shown in orange-, red- and purple-colour lines, respectively. Vertical lines show the medians of the samples.}    
   \label{fig:density}
\end{figure}
\subsection{Local density}


To complement the environmental context, we quantify local density with the 5th–nearest–neighbour estimator, $\Sigma_5 = 5/(\pi r_5^2)$, where $r_5$ is the projected distance to the 5th nearest tracer galaxy. Tracers are drawn from a volume-limited sample with $M_r \le -20.5$, ensuring homogeneity over our redshift range and mitigating redshift-dependent incompleteness, this is the threshold follows standard SDSS practice (e.g. \citealt{Balogh2004, 2004Baldry}), lies close to $L^\star$ in $r$ (\citealt{2003Blanton}), and has been adopted in S-PLUS applications as well (\citealt{2024Cerdosino}). For CGs, we identify neighbours around the CG centre and exclude CG members from the tracer set. For surrounding-group galaxies, $r_5$ is measured around each galaxy using the same tracer definition.

Figure \ref{fig:density} shows the distributions of projected local densities ($\log(\Sigma_5)$) for three samples: (1) isolated CGs (orange), (2) surrounding group galaxies (purple), and (3) non-isolated CGs (red). The median values for the local densities of these are respectively $\sim -1.87$, $\sim-1.79$ and $\sim-1.83$. The distributions reveal a modest environmental trend, with slight differences in the median local densities among the three samples. Focusing on the distribution in the range of log⁡($\Sigma_5$) between -3.2 and -0.5, we observe that isolated CGs peak at the lowest density of the three (log⁡($\Sigma_5$)$\sim-2.3$), while non-isolated CGs peak at the highest density (log⁡($\Sigma_5$)$\sim-1.5$), a difference that is expected by construction. Meanwhile, the surrounding group galaxies exhibit a peak right in between the two distributions. 

This visual modest environmental trend is supported by a Kolmogorov–Smirnov (K–S) statistical test, which confirms significant differences in the $\Sigma_5$ distributions between isolated and non-isolated CGs (p-value $\sim 6\times10^{-4}$), as well as between surrounding group galaxies and non-isolated CGs (p-value $\sim 2\times10^{-3}$). In contrast, the distributions of isolated CGs and surrounding group galaxies are statistically consistent (p-value = 0.27), suggesting that we can not reject the null hypothesis, therefore the environment of isolated CGs and that of the surrounding groups galaxies are not significantly different in terms of local projected density.



We also observe a peak at very low local densities (below -3.2) for non-isolated CGs and surrounding group galaxies. These correspond to CGs embedded in the outermost regions of major structures or within loose groups. In the case of a surrounding group galaxy, the low density may indicate that the galaxy lives in a sparse loose group.

\subsection{Velocity dispersion and $R_{200}$}

To estimate the velocity dispersion of major structures, we adopted the gapper estimator method introduced by \cite{1990AJ....100...32B}. This approach is more robust than the standard deviation for small samples, as demonstrated by the authors. Consequently, we used the $\sigma_{\rm{G}}$ values available for the 102 major structures. The rest-frame velocity dispersion is computed as:

\begin{equation}
\sigma_{\rm{G}} = \frac{\sqrt{\pi}}{(1+z_g)N(N-1)} \sum_{i=1}^{N-1} w_i g_i,
\end{equation}

where $z_g$ corresponds to the redshift of the major structures, and $N$ is the number of member galaxies. Each term $w_i$ is a Gaussian weight given by $w_i = i(N-i)$, and $g_i$ represents the velocity gap between adjacent galaxies, defined as, $g_i = V_{i+1} - V_i$, with $V_i$ being the radial velocity of the $i$-th galaxy in the group.

To derive the characteristic group radius of the major structures, we adopt the method described by \cite{2024Epinat} and references therein (e.g., \citealt{2012Lemaux}, \citealt{1997Carlberg}, \citealt{2006Biviano}). We compute $R_{200}$, the radius within which the mean density is 200 times the critical density of the Universe at $z_{\rm{g}}$, using the expression:

\begin{equation}
R_{200} = \frac{\sqrt(3)\sigma_{\rm{G}}}{10H(z_{\rm{g}})},
\end{equation}

where $H(z_{\rm{g}})$ is the Hubble parameter at the redshift of the major structure.

\begin{figure}
    \centering
    \includegraphics[scale=0.4]{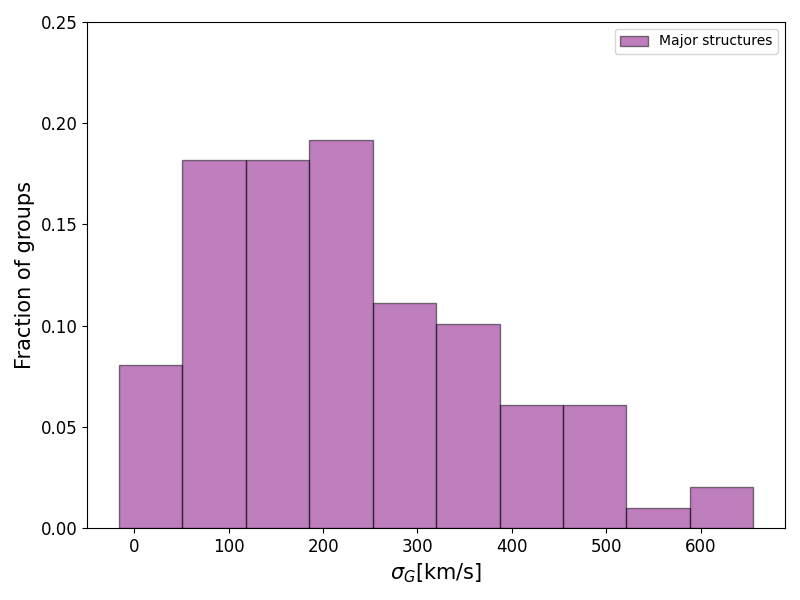}
    \includegraphics[scale=0.4]{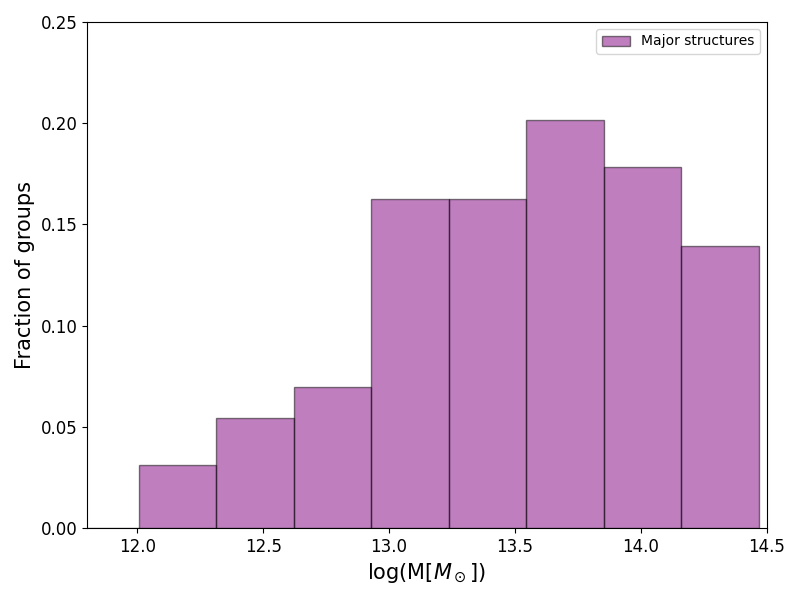}
    \caption{Host-structure properties. Left: velocity dispersion estimated in this work. Right: halo mass from \citet{2007Yang}. The broad ranges arise because some hosts are small substructures with few members, whereas others contain $>50$ galaxies, leading to higher $\sigma_G$ and larger $M_{halo}$.}    
   \label{fig:host}
\end{figure}

In Figure \ref{fig:host}, we present the distribution of the velocity dispersion estimated by us (left panel) for the major structures, and the halo mass estimated by \cite{2007Yang} (right panel) for the same systems. In both plots, we observe that the major structures span a wide range of masses and velocity dispersions. This is because, within the larger structures, we find substructures composed of a small number of galaxies — for instance, systems with only six members, three belonging to a CG and three to the surrounding sample — which explains why some major structures show relatively low velocity dispersions and halo masses. On the other hand, there are also major structures containing more than 50 galaxies, which results in higher velocity dispersions and larger halo masses.

\section{Results}
\label{sec:results}

\subsection{Wavelength dependence of structural parameters}
\label{sec:sub_n_re_lam}

\begin{figure}
    \centering
     \includegraphics[scale=0.25]{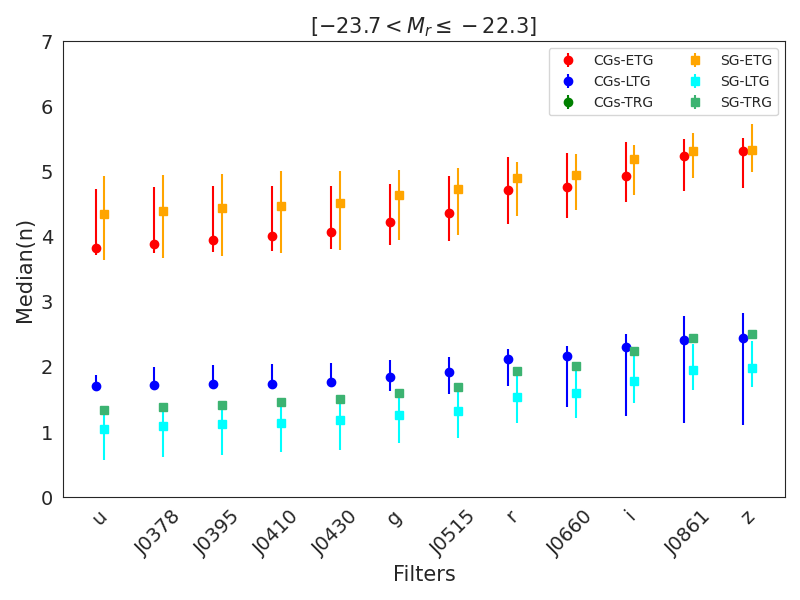}
    \includegraphics[scale=0.25]{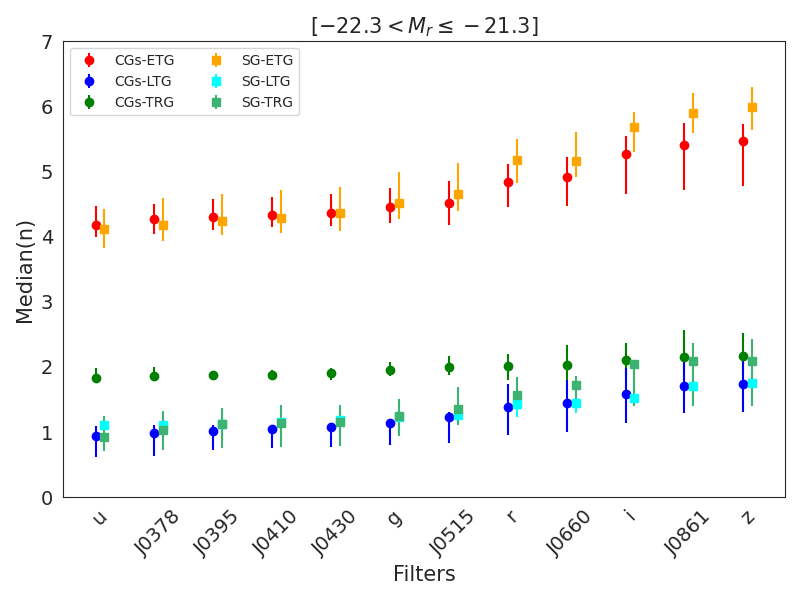}
    \includegraphics[scale=0.25]{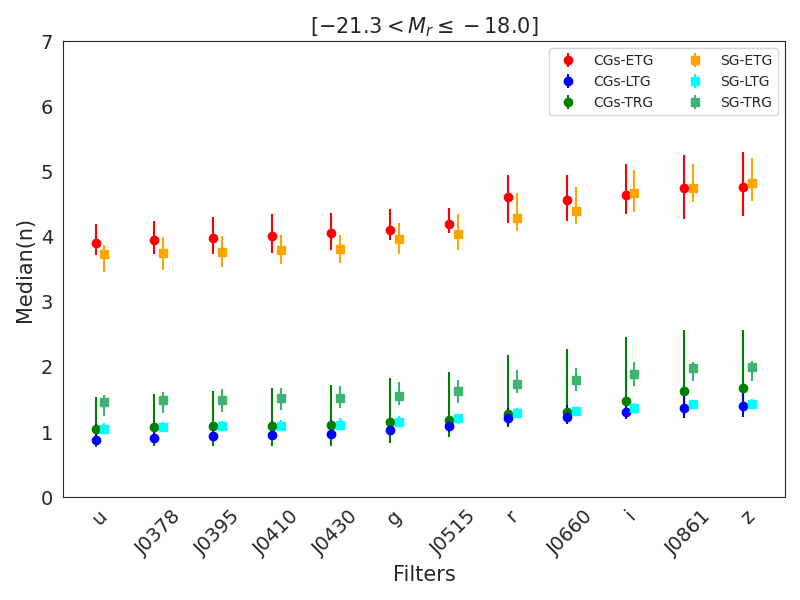}
    \includegraphics[scale=0.25]{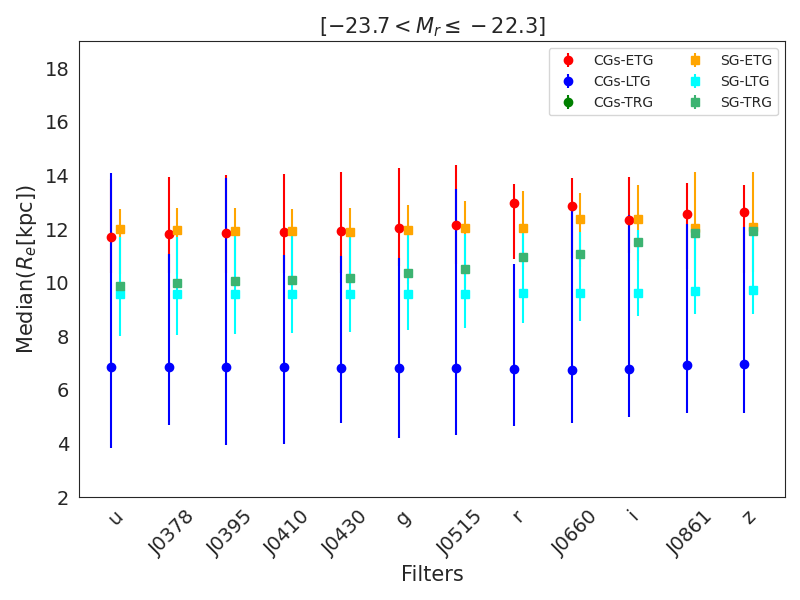}
    \includegraphics[scale=0.25]{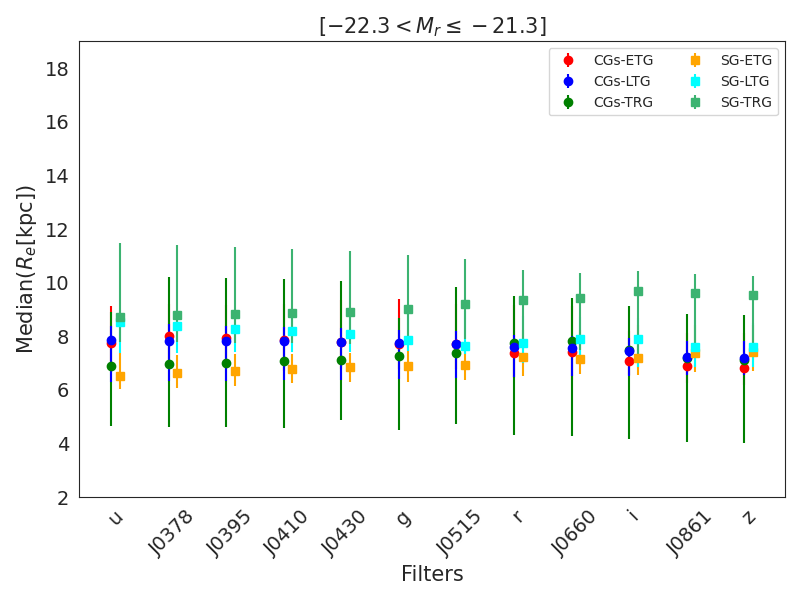}
    \includegraphics[scale=0.25]{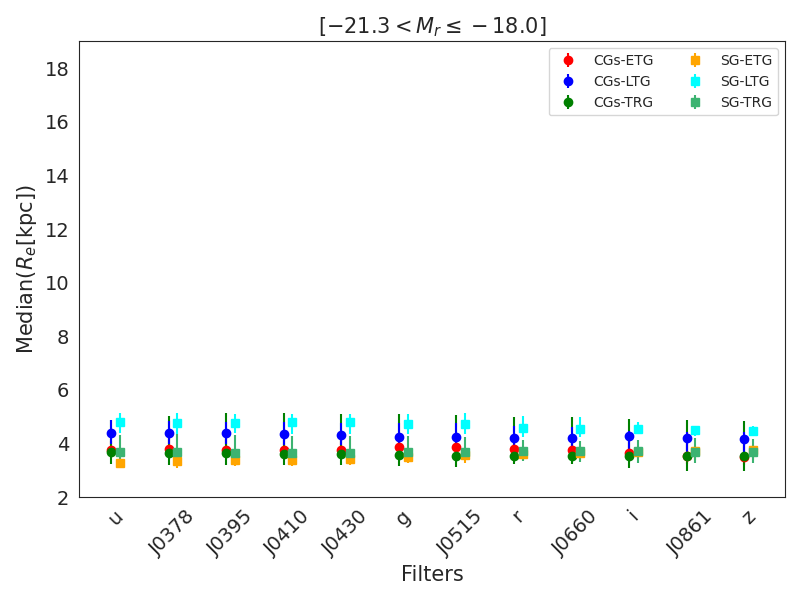}
    \caption{Top panels: Median Sérsic index ($n$) as a function of wavelength for ETGs (red/orange), LTGs (blue/cyan), and TGs (green/light green) in non-isolated compact groups (circles) and their surrounding groups (squares). Bottom panels: Median effective radius ($R_e$) as a function of wavelength for the same subsamples. The three panels correspond to decreasing luminosity bins from left to right. Error bars show the 90\% confidence intervals estimated by bootstrapping.}
    \label{fig:n_median}
\end{figure}

Figure~\ref{fig:n_median} shows the median Sérsic index ($n$, top panels) and the median effective radius ($R_e$, bottom panels) as a function of wavelength for galaxies in non-isolated CGs (circles) and their surrounding group galaxies (SGs; squares). ETGs, LTGs, and TGs are shown in red/orange, blue/cyan, and green/light green, respectively. Uncertainties were estimated using bootstrapping with a 90\% confidence interval. The three panels correspond to decreasing luminosity bins, from left to right: $-23.7 < M_r \leq -22.3$, $-22.3 < M_r \leq -21.3$, and $-21.3 < M_r \leq -18.0$. These bins were chosen to ensure comparable numbers of TGs between environments, allowing a reliable morphological comparison at least in the intermediate and faint absolute-magnitude bins for the non-isolated CGs. In the brightest bin, however, there is only one TG in the surrounding sample, so the comparison is not statistically meaningful.

In both environments, the Sérsic index $n$ increases with wavelength for all morphological types, consistent with previous studies (\citealt{2010LaBarbera}; \citealt{Kelvin2012}; \citealt{2014vulcani}; \citealt{2021Lima-Dias}; \citealt{2023Montaguth}). The effect is strongest for LTGs, which display the largest variation in $n$ across filters due to their composite bulge–disc structures \citep{2014vulcani}. ETGs show smaller but consistent increases of $n$, as expected for systems dominated by old stellar populations with mild colour gradients. TGs exhibit intermediate behaviour between ETGs and LTGs, with moderate growth of $n$ from blue to red filters.

The effective radius, varies with wavelength. Among bright ETGs, $R_e$ tends to increase towards redder filters in both environments. At intermediate and faint luminosities, an environmental dependence emerges: in non-isolated CGs $R_e$ decreases with wavelength, whereas in surrounding group galaxies it increases. LTGs generally exhibit flat or mildly declining $R_e(\lambda)$ relations, as expected for disc–dominated systems, with the exception of the bright bin, where a slight increase with wavelength is observed. TGs show either a gentle rise towards redder filters or an approximately wavelength–independent $R_e$, depending on luminosity. Taken together, these trends support the view that wavelength–dependent structural variations are closely linked to internal stellar–population gradients \citep{2010LaBarbera}.

\begin{figure}
    \centering
    \includegraphics[width=0.7\linewidth]{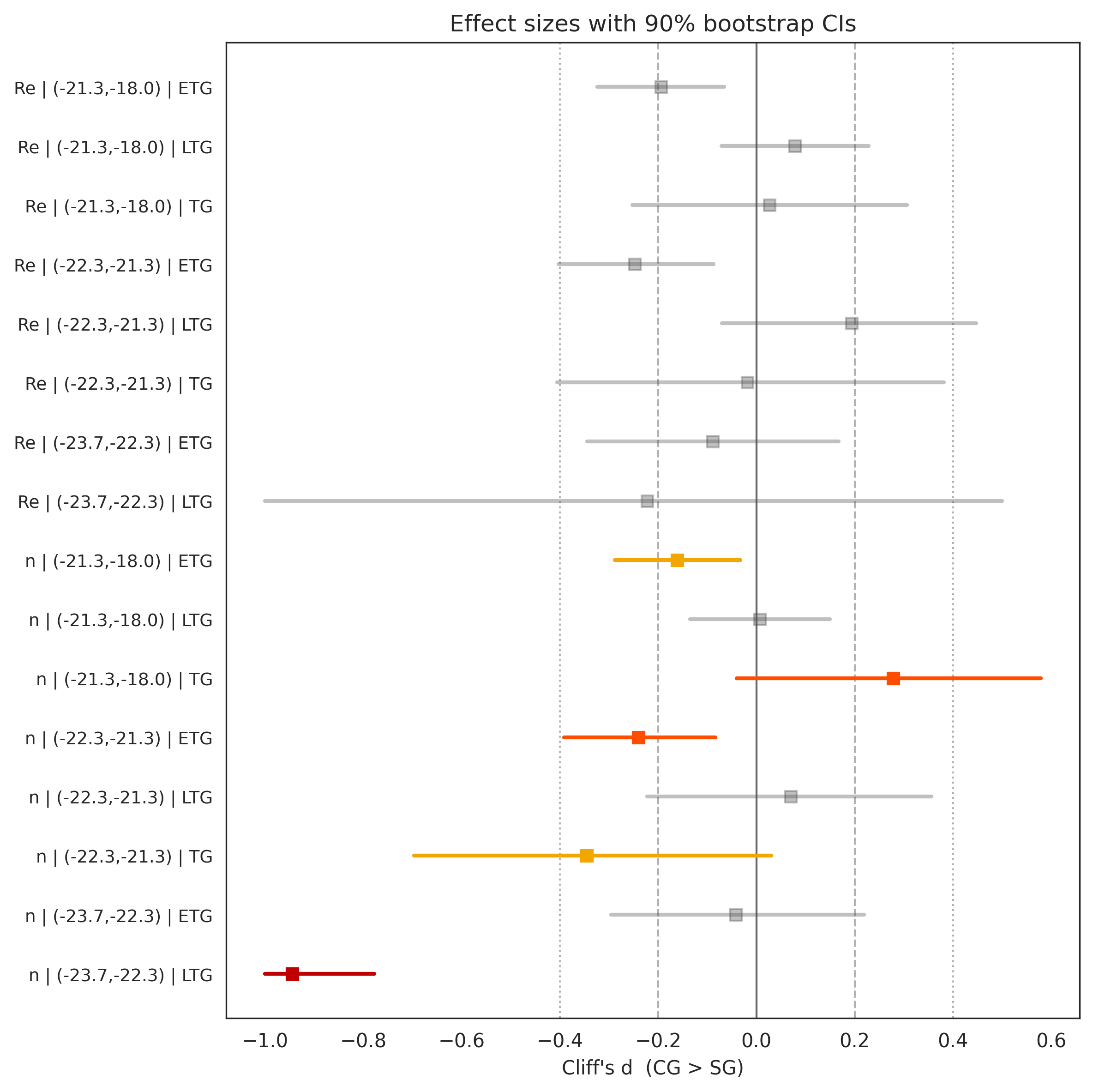}
    \caption{Cliff’s delta effect sizes ($\delta$) comparing the slopes of $n(\lambda)$ and $R_e(\lambda)$ between galaxies in non-isolated compact groups (CGs) and their surrounding groups (SGs). Negative values indicate steeper structural gradients in CGs. Horizontal bars denote 90\% bootstrap confidence intervals. Colours indicate the significance level from the permutation test: red ($p_{\mathrm{perm}}<0.01$), orange ($p_{\mathrm{perm}}<0.05$), and amber ($p_{\mathrm{perm}}<0.10$).}
    \label{fig:forest_cliffs}
\end{figure}

To quantify whether these wavelength-dependent variations differ between environments, we measured for each galaxy the slope of $n(\lambda)$ and $R_e(\lambda)$ as a function of $\log( \lambda)$. We then compared the resulting slope distributions between non-isolated CGs and surrounding groups using a non-parametric permutation test and the Cliff’s delta effect size (Figure~\ref{fig:forest_cliffs}). The permutation test ($p_{\mathrm{perm}}$) evaluates the probability that the observed difference arises by chance, while Cliff’s delta ($\delta$) measures the direction and magnitude of the effect, being negative when non-isolated CG galaxies exhibit steeper or more compact structural gradients than surrounding groups galaxies.

The resulting forest plot (Figure~\ref{fig:forest_cliffs}) shows that, for ETGs, the slopes of both $n(\lambda)$ and $R_e(\lambda)$ are systematically steeper (more negative) in non–isolated CGs than in surrounding groups at intermediate and faint luminosities. For $n(\lambda)$ the effect is small–to–moderate (Cliff’s $\delta\simeq -0.23$ and $-0.16$, with $p_{\mathrm{perm}}=0.033$ and $0.067$, respectively), while $R_e(\lambda)$ exhibits weaker negative shifts (Cliff’s $\delta\simeq -0.23$ and $-0.19$; $p_{\mathrm{perm}}=0.169$ and $0.106$). In the brightest bin, ETG differences are small and not statistically significant, for instance, $n(\lambda)$ has Cliff’s $\delta\simeq -0.06$ with $p_{\mathrm{perm}}=0.221$, and $R_e(\lambda)$ has Cliff’s $\delta\simeq -0.10$ with $p_{\mathrm{perm}}=0.516$.
LTGs remain largely consistent across environments in all luminosity bins (all $|\delta|\lesssim 0.22$, $p_{\mathrm{perm}}\gtrsim 0.3$), with a single exception: in the brightest bin, $n(\lambda)$ differs markedly ($p_{\mathrm{perm}}\simeq 0.0038$) and shows a very large negative effect (Cliff’s $\delta\simeq -0.94$), although the non–isolated CG sample there is small ($n_{\mathrm{CG}}=4$), so this result warrants caution.
TGs reveal a mixed behaviour: at intermediate luminosities, $n(\lambda)$ is modestly steeper in non–isolated CGs (Cliff’s $\delta\simeq -0.35$, $p_{\mathrm{perm}}=0.096$), whereas at faint luminosities the trend reverses and surrounding groups show steeper $n(\lambda)$ (Cliff’s $\delta\simeq +0.28$, $p_{\mathrm{perm}}=0.020$). For $R_e(\lambda)$, TGs display no significant environmental differences in any bin ($p_{\mathrm{perm}}\simeq 0.69$ and $0.48$ at intermediate and faint luminosities, respectively), and in the brightest bin no comparison is possible because there are no TGs in the non–isolated CG sample.

\subsection{The effective radius vs. Sérsic index plane}

In \cite{2023Montaguth}, we identified a bimodal distribution in the $R_e-n$ plane for transition galaxies in CGs, where one population, not found in field galaxies, consists of peculiar galaxies that are more compact (higher $n$) and smaller (lower $R_e$) compared to field galaxies. Moreover, in \cite{2024Montaguth}, we found that transition galaxies in non-isolated CGs exhibit a bimodal distribution in the $R_e$–$n$ plane, with Sérsic indices smoothly increasing towards higher values above 1.5, suggesting that the majority of these galaxies have already undergone morphological transformation. In contrast, transition galaxies in isolated CGs typically have n$<1.75$, suggesting that they are at an earlier stage of morphological evolution. Therefore, exploring this plane could help us identify potential morphological differences between transition galaxies in non-isolated CGs and their counterparts in surrounding group galaxies.

\begin{figure}
    \centering
    \includegraphics[scale=0.4]{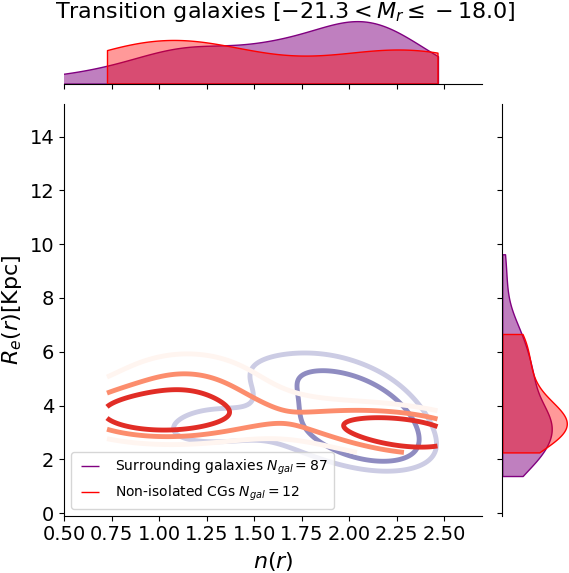}
    \includegraphics[scale=0.4]{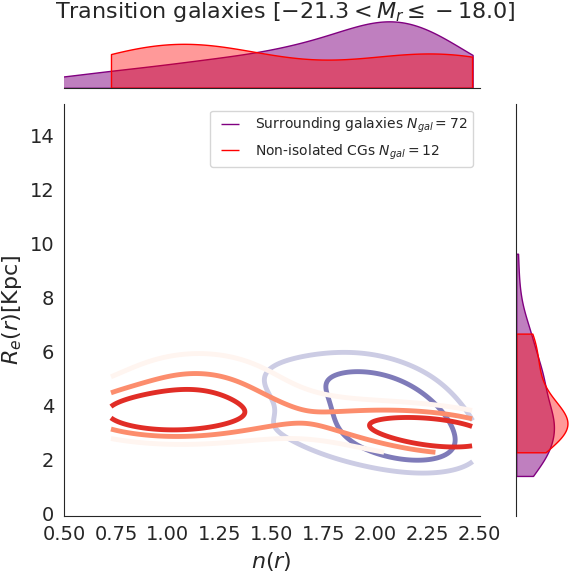}
    \caption{Contours of effective radius as a function of Sérsic index in the $r$-band for transition galaxies, for non-isolated CGs in red contours, and for the surrounding group galaxies in purple. The distribution of each parameter is shown in the margin of each plot. The left panel shows galaxies with intermediate absolute magnitudes (-22.3$< M_r \leq$ -21.3) in the $r$-band, while the right panel displays those in the low-luminosity bin (-21.3$< M_r \leq$ -18.0).}
   \label{fig:re_n_TG}
\end{figure}

In Figure \ref{fig:re_n_TG}, we present $R_e$ as a function of $n$ in the $r$-band for transition galaxies. The left panel shows galaxies with intermediate absolute magnitudes in the $r$-band, while the right panel displays those in the faint magnitude bin. We found that there are no transition galaxies in the brightest absolute magnitude bin. Red contours represent the density distribution of transition galaxies in non-isolated CGs, whereas purple contours correspond to their counterparts in surrounding groups. In the intermediate absolute magnitude bin, the contours for transition galaxies in non-isolated CGs are located at $n$ values above 1.5, while the contours for the surrounding group galaxies are more widely spread, ranging from 0.5 to 2. Additionally, transition galaxies in non-isolated CGs exhibit smaller $R_e$ values compared to those in the surrounding group galaxies. These trends are also evident in Figures \ref{fig:n_median}, where we found that the median value for $n$ was higher, while $R_e$ was slightly lower for non-isolated CG galaxies compared to those for surrounding group galaxies. Interestingly, the majority of transition galaxies in non-isolated CGs show $n$ values greater than 2, which corresponds to the peculiar galaxy population defined in \cite{2023Montaguth}. In contrast, surrounding group galaxies reach a higher density around $n\sim1.5$.

For galaxies in the faint absolute magnitude bin, we observe a bimodal distribution for transition galaxies in non-isolated CGs, similar to the bimodality observed for all transition galaxies in \cite{2023Montaguth}, while the contours for surrounding group galaxies show a unimodal distribution, concentrated in one of the regions corresponding to the bimodality observed for the more compact galaxies (with $n>1.7$) in non-isolated CGs.

To assess whether the visual differences previously described in the $R_e-n$ distributions are statistically significant, we applied a bootstrap analysis based on the Mahalanobis distance \citep{1936Mahalanobis}. This multivariate statistical test evaluates whether two groups differ in a multidimensional space by measuring the distance between their central tendencies (medians, in this case), while accounting for the scale and correlation of the involved variables, in this case, $n$ and $R_e$. In the intermediate absolute magnitude bin, the Mahalanobis distance between the medians of galaxies in non-isolated CGs and those in surrounding groups was 1.08, with a 95\% confidence interval of [0.50, 2.71], based on bootstrap resamplings. Similarly, in the faint bin, the observed distance was 1.12, with a confidence interval of [0.22, 1.64]. In both cases, the confidence intervals do not include zero, indicating that the structural differences between the two samples are statistically significant. These results confirm that the trends observed in $R_e-n$ diagrams reflect real differences in the structural properties of transition galaxies in non-isolated CGs compared to those in surrounding groups. While these results indicate statistically significant structural differences, we note that the sample size is relatively small. Therefore, caution is warranted when generalising these findings, and further confirmation with larger datasets would be valuable.

\subsection{Stellar mass dependence of quenched galaxies and morphological fractions}
\label{sec:ssfr}

\begin{figure}
    \centering
    \includegraphics[scale=0.34]{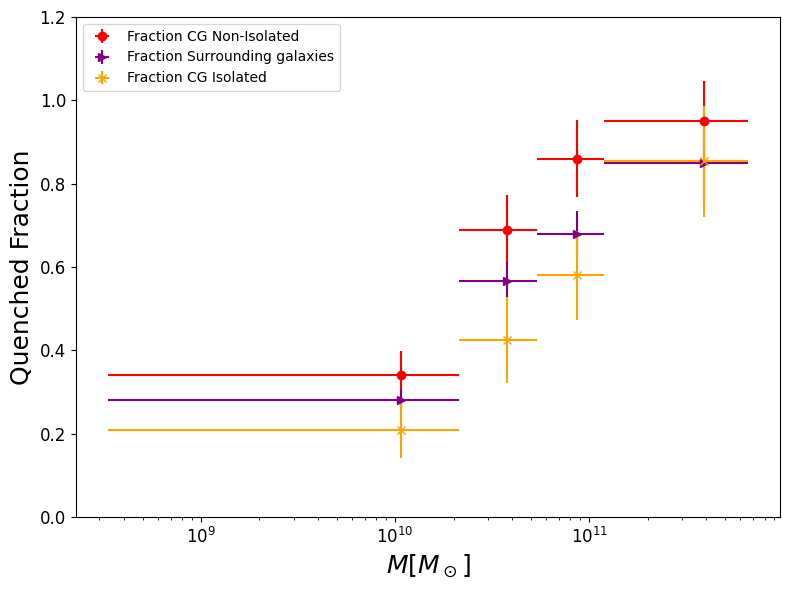}
    \includegraphics[scale=0.34]{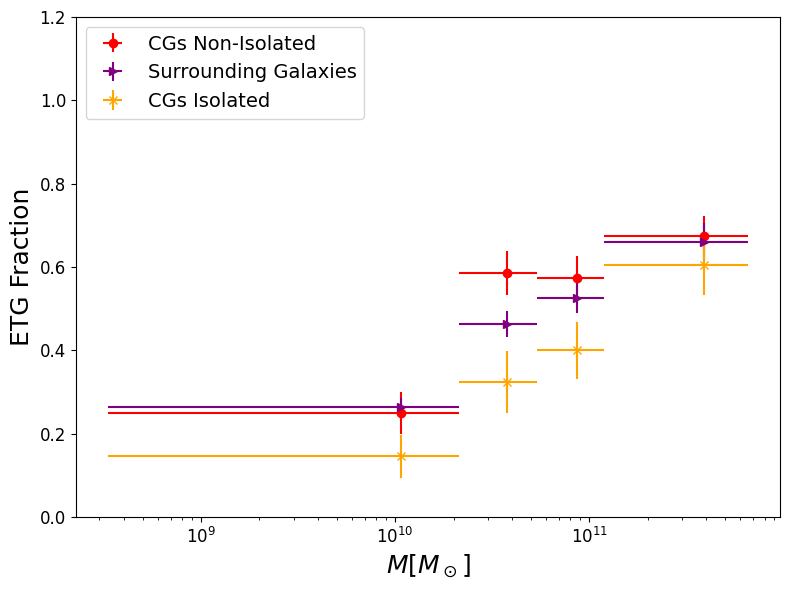}
    \caption{In the top panel, we present the quenched galaxy fraction as a function of stellar mass, while in the bottom panel, we show the fraction of ETGs instead of that of quenched galaxies. In both panels, galaxies in non-isolated CG are represented in red, surrounding group galaxies in purple, and isolated CGs in orange. The fraction of galaxies is calculated in five stellar mass bins. The error bars along the x-axis indicate the width of each mass bin, while the y-axis error bars represent the Poisson counting error.}
   \label{fig:que_frac}
\end{figure}

In Figure \ref{fig:que_frac}, the fraction of quenched galaxies (upper panel) and ETGs (bottom panel)  across different stellar mass bins is shown. A galaxy is considered quenched if $log(sSFR[yr^{-1}]) \leq -11$, following the criterion proposed by \cite{2006Weinmann}. The fraction of galaxies is estimated as the number of quenched galaxies, or ETGs, normalised by the total number of galaxies within the bin. Additionally, to facilitate the comparison, we added the isolated CGs analyzed in \cite{2024Montaguth} to the plots, which are shown in orange.

We observe that the fraction of quenched galaxies is higher in non-isolated CGs than in surrounding group galaxies and isolated CGs, especially in the higher mass bins. The same behaviour is seen for the fraction of ETG galaxies. These plots suggest that non-isolated CGs may promote morphological transformations, and the quenching process, more strongly than surrounding group galaxies and isolated CGs. This result is in agreement with the findings of \cite{2023Taverna}, who found that the fractions of red galaxies and ETGs are higher in CGs compared to the surrounding group galaxies. 

\begin{figure*}
    \centering
    \includegraphics[width=0.33\textwidth]{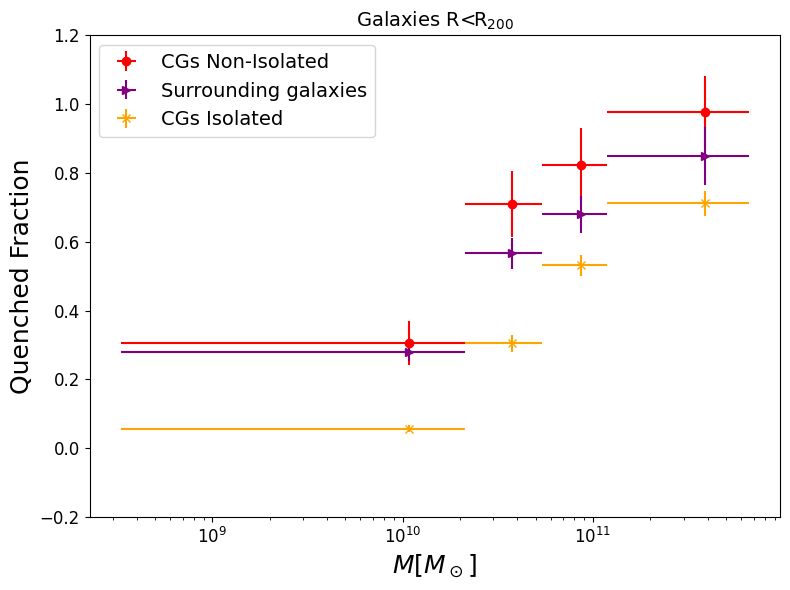}
    \includegraphics[width=0.33\textwidth]{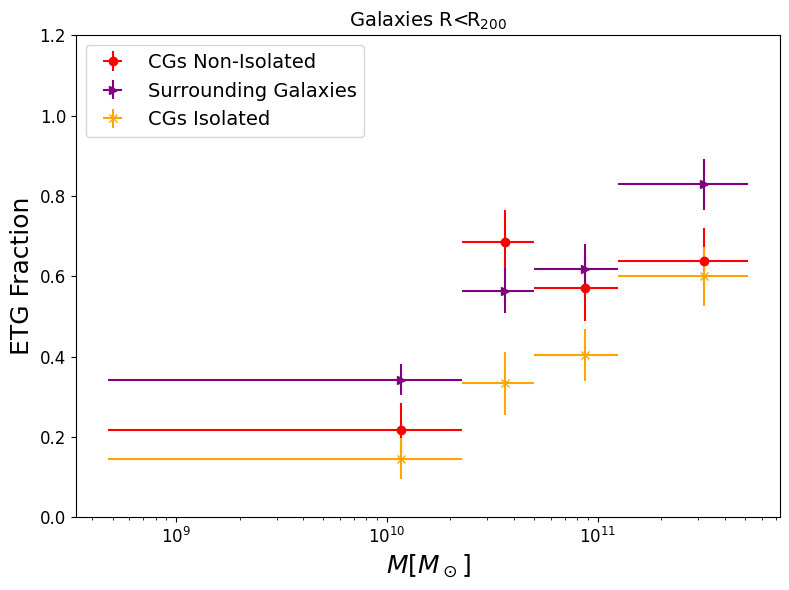}\\
    \includegraphics[width=0.33\textwidth]{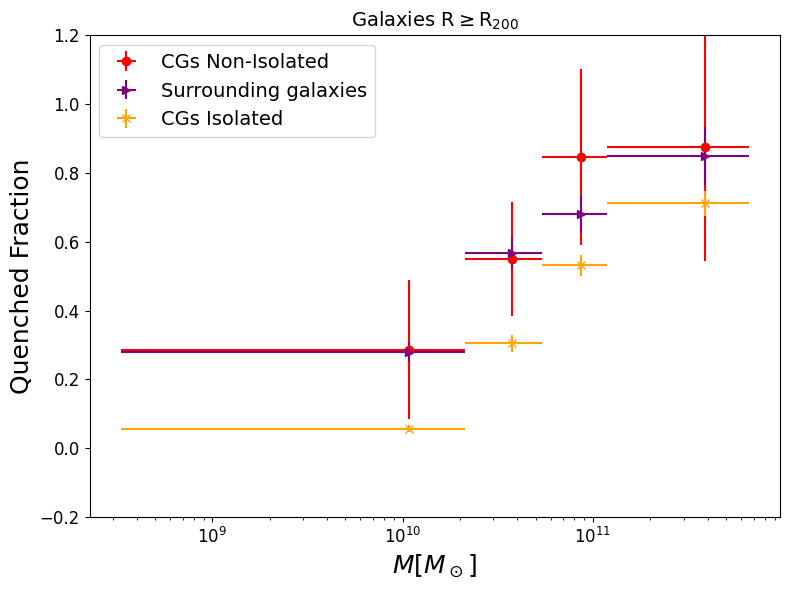}
    \includegraphics[width=0.33\textwidth]{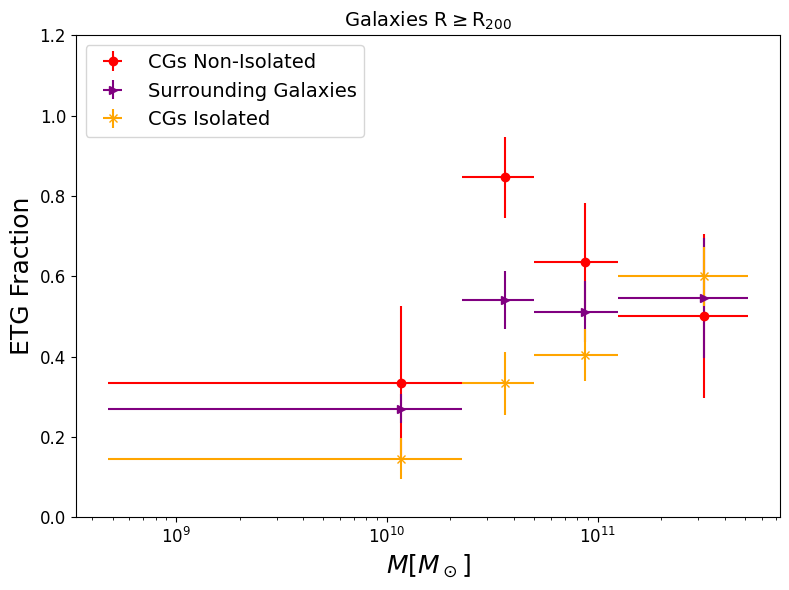}
    \caption{Quenched galaxy (left panels) and ETG (right panels) fraction as a function of stellar mass. Galaxies in non-isolated CGs are represented in red, surrounding group galaxies in purple, and isolated CGs in orange. In the top panels, we show surrounding group and non-isolated CG galaxies within the $R_{200}$ of the major structure where the CGs are embedded, while in the bottom panels, we show the fraction of galaxies located outside $R_{200}$. In the case of isolated CGs, we present all the galaxies for comparison purposes. The fraction of galaxies is calculated in five stellar mass bins. The error bars along the x-axis indicate the width of each mass bin, while those along the y-axis represent the Poisson counting error.}
    \label{fig:que_frac_dis}
\end{figure*}

Figure~\ref{fig:que_frac_dis} shows, similarly to Figure~\ref{fig:que_frac}, the fraction of quenched galaxies and ETGs as a function of stellar mass. In this case, the top panels present galaxies located within the $R_{200}$ of the major structure in which the CGs are embedded, while the bottom panels show galaxies located beyond $R_{200}$. The mass bins are constructed to contain the same number of galaxies as in the non-isolated CG sample. 

For quenched galaxies within $R_{200}$ (top left panel), the fraction is generally higher for CG galaxies compared to their surrounding groups, except for the lowest stellar mass bin. The quenched fraction in both non-isolated CGs and surrounding groups within $R_{200}$ is also higher than that of isolated CGs. For galaxies located beyond $R_{200}$ (bottom left panel), the quenched fraction remains slightly higher for CGs compared to surrounding groups, but their values are consistent within the error bars.

Regarding the ETG fraction (right panels), for CGs located in the inner regions of the host structures (top right panel), the fraction is generally lower than in their surrounding groups, except in the $log(M[M_{\odot}]) = 10.3$–$10.7$ bin, where CGs exhibit a higher ETG fraction. For non-isolated CGs located outside of $R_{200}$ (bottom right panel), we find that the ETG fraction is higher in non-isolated CGs compared to galaxies in the surrounding group sample, and isolated CGs, except in the most massive bin, where all three environments present similar values. Indeed, when comparing the fraction of quenched galaxies with that of ETGs, we observe that, for non-isolated CGs located beyond $R_{200}$, the ETG fraction is higher in the two least massive bins.

\begin{figure*}
    \centering
    \includegraphics[width=0.33\textwidth]{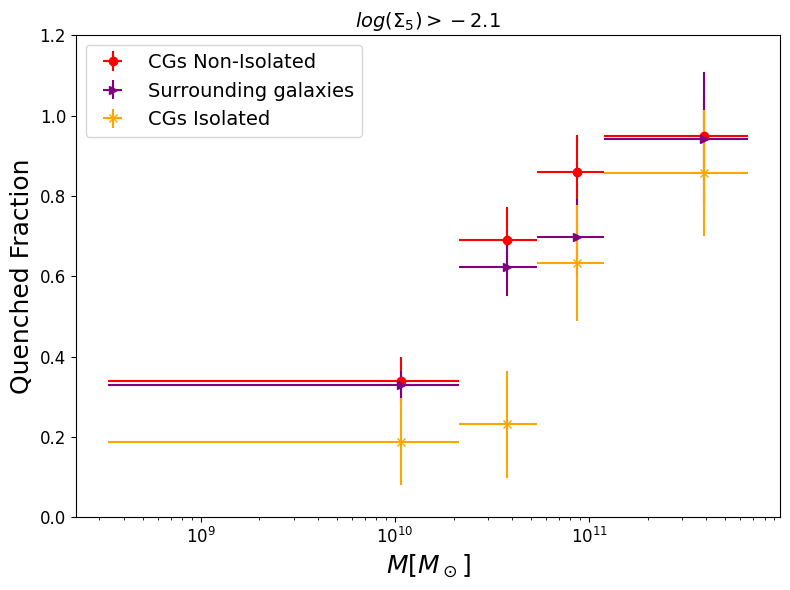}
    \includegraphics[width=0.33\textwidth]{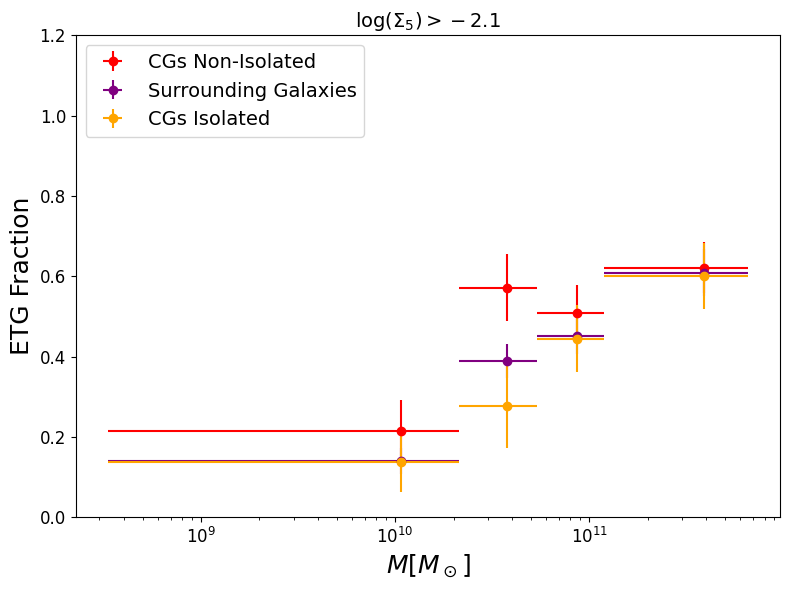}\\
    \includegraphics[width=0.33\textwidth]{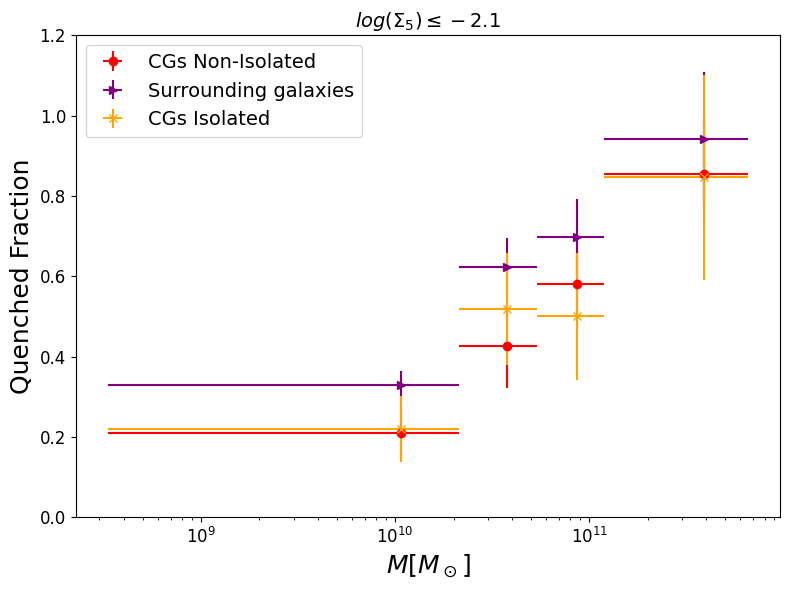}
    \includegraphics[width=0.33\textwidth]{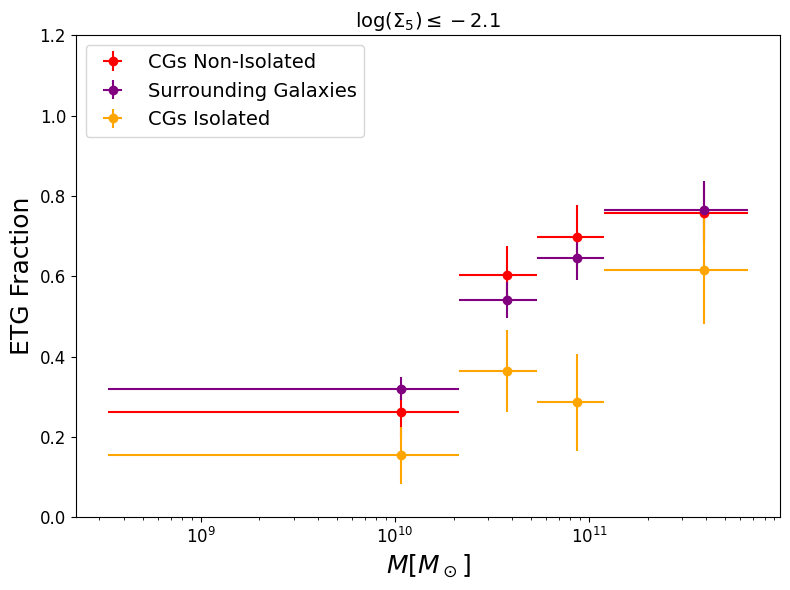}
    \caption{Quenched galaxy (left panels) and ETG (right panels) fraction as a function of stellar mass. Galaxies in non-isolated CGs are represented in red, surrounding group galaxies in purple, and orange isolated CGs. In the top panels, we show surrounding group galaxies, non-isolated CG, and isolated CGs with log($\Sigma_5$)$<$-2.1, while in the bottom panels, we show the fraction of galaxies with log($\Sigma_5$)$<$-2.1. The fraction of galaxies is calculated in five stellar mass bins. The error bars along the x-axis indicate the width of each mass bin, while those along the y-axis represent the Poisson counting error.}
    \label{fig:que_frac_dens}
\end{figure*}

Figure~\ref{fig:que_frac_dens} presents the fraction of quenched galaxies (left panels) and ETGs (right panels) as a function of stellar mass, now split by local density: high-density ($\log\Sigma_5 > -2.1$, top panels) and low-density ($\log\Sigma_5 \leq -2.1$, bottom panels). We chose this threshold because it corresponds to the median value of the distribution when considering all three samples: surrounding galaxies, non-isolated CGs, and isolated CGs. In both density regimes, quenched fractions are systematically higher in non-isolated CGs compared to isolated CGs, particularly at intermediate and high stellar masses ($log(M[M_{\odot}]) \gtrsim 10$). At low densities (bottom left), the quenched fraction in surrounding groups approaches that of CGs, with differences reduced relative to the high-density case. In terms of morphology (right panels), the ETG fraction shows greater dispersion, especially at intermediate masses in low-density environments (bottom right). In high-density regions (top right), CGs exhibit lower ETG fractions than surrounding groups, while isolated CGs maintain the lowest ETG fractions across all mass bins.

\subsection{Dynamics of CGs Embedded in Clusters}
\label{sec:dyn}

\begin{figure*}
    \centering
    \includegraphics[width=0.35\textwidth]{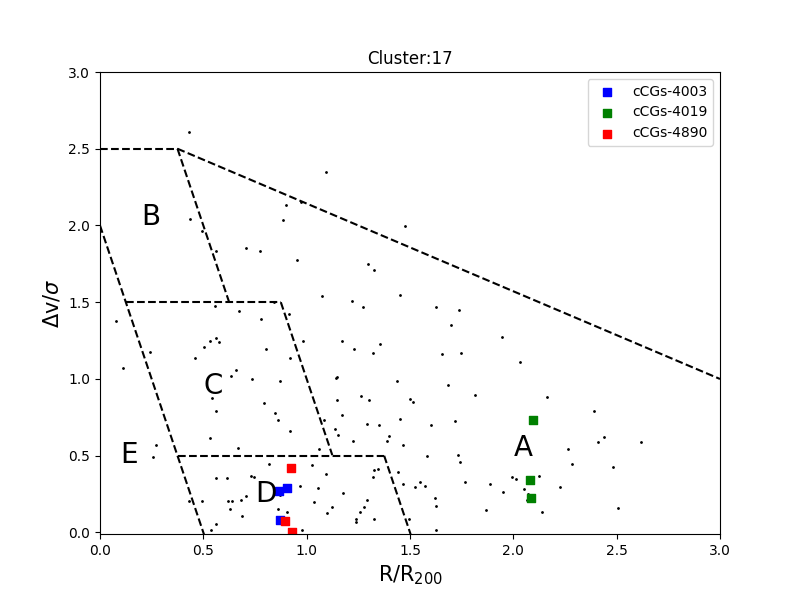}
    \includegraphics[width=0.35\textwidth]{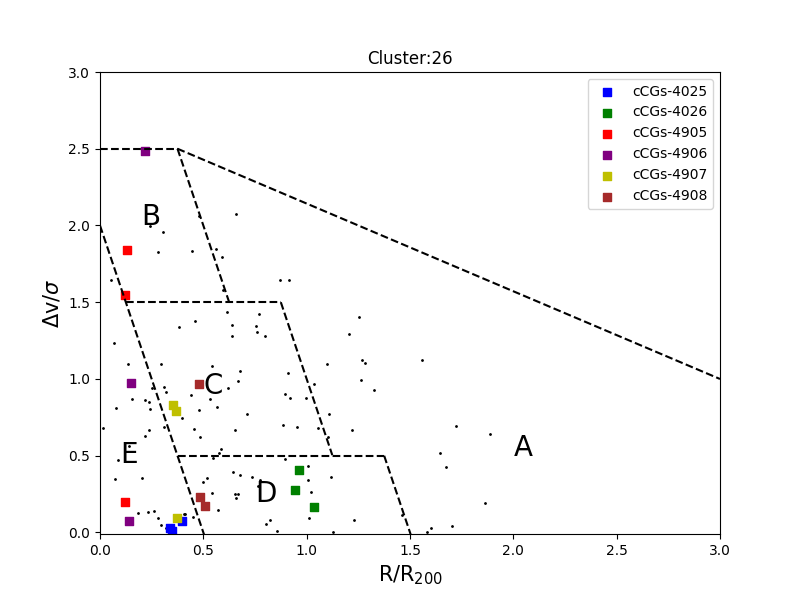}
    \includegraphics[width=0.35\textwidth]{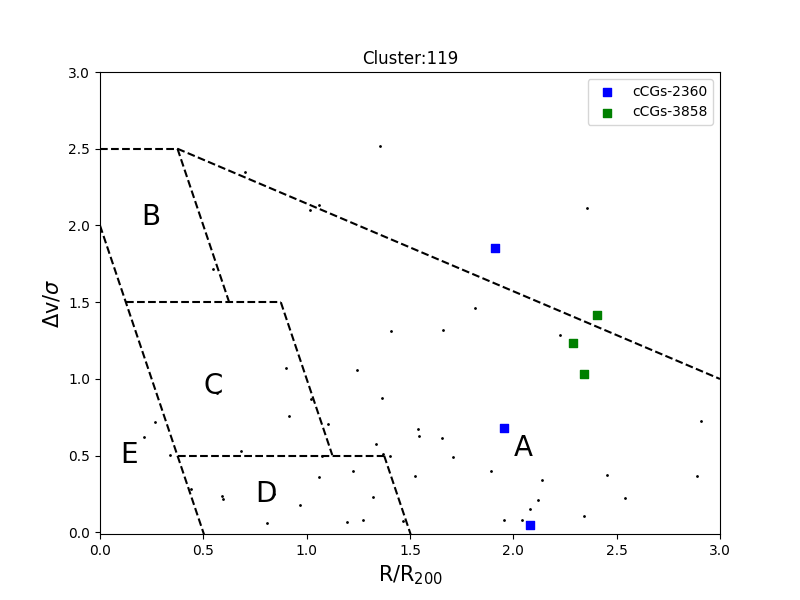}
    \includegraphics[width=0.35\textwidth]{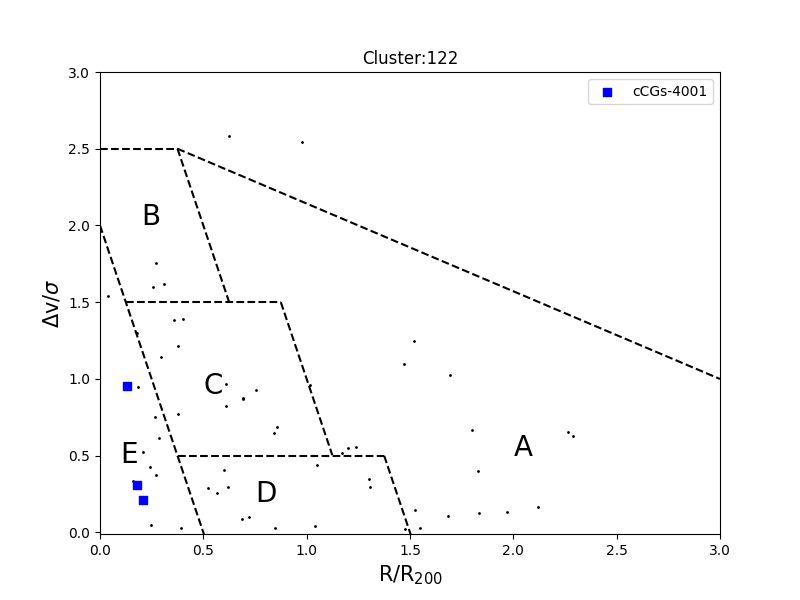}
    \caption{Projected phase-space diagrams (PPSD) for four clusters's, showing the line-of-sight velocity difference of each galaxy with respect to the mean cluster velocity relative to the observer, normalised by the cluster's velocity dispersion, as a function of the projected radius from the cluster centre, normalised by the cluster's $R_{200}$. The coloured squares represent galaxies identified as part of a CG, with different colours corresponding to each CG, as indicated in the legend. The black dots denote the remaining galaxies in the clusters. The regions in the PPSD, as defined by \cite{Rhee_2017}, correspond to different stages of infall.}
    \label{fig:PSD}
\end{figure*}

This projected phase–space analysis is restricted to the subset of CGs embedded in hosts with at least 50 spectroscopic members, for which $(R_{200},\sigma$ are reliable. In our data this corresponds to $N_{\rm cl}=4$ clusters hosting $N_{\rm CG}^{\rm PPSD}=12$ CGs. We do not extrapolate the trends reported here to the remaining systems. The $N\ge 50$–member threshold is an operational requirement to obtain stable estimates of $\sigma$ and $R_{200}$; for poorer systems these quantities become too uncertain for a meaningful phase–space classification.

From a dynamical standpoint, these embedded CGs can be examined to determine whether they are currently being accreted onto the host clusters or are already gravitationally bound to them. Moreover, by employing the projected phase-space diagram (PPSD; \citealt{Hernández-Fernández2014}, \citealt{Jaffe2015}), we can assess whether these CGs are physically bound systems or merely the result of chance alignments.

In Figure \ref{fig:PSD}, we present the PPSDs for these four clusters. The plots show the line-of-sight velocity difference of each galaxy relative to the cluster, normalised by the cluster’s velocity dispersion, as a function of the projected radius of each galaxy from the cluster centre, normalised by the cluster’s $R_{200}$. The coloured squares represent galaxies identified as part of a CG, with each CG colour-coded as indicated in the legend of each plot. The black dots represent the remaining galaxies within the clusters, that is, the surrounding group galaxies. The regions in the PPSD are defined by \cite{Rhee_2017} through simulations, which established five zones to statistically infer the infall time, that is, the time since a galaxy entered the cluster. Galaxies in zone A are likely first infallers, while those in zone B are recent infallers. Zone D contains intermediate infallers, and zone C is a mixture of zones B and D. Finally, galaxies in zone E are considered ancient infallers.

In the case of cluster number 17 (see the top-left plot in Figure \ref{fig:PSD}), there are three CGs identified according to the catalogues of \cite{zheng2020compact}. Two of these groups are located in the intermediate infallers region, while one is situated in the first infallers region. For cluster number 26, we observe 6 CGs. None of the CGs fall within zone A, of first infallers. In addition, we noticed distinct behaviours among the CGs. For instance, in the cases of CGs-4905 and CGs-4906, the galaxies share similar projected radii but exhibit significant differences in their line-of-sight velocities relative to the cluster centre. This discrepancy suggests that, although the galaxies appear to form a group in projection, they are not physically bound, as indicated by the lack of coherence in $\Delta v/\sigma$. On the other hand, for CGs-4908 and CGs-4907, where two galaxies are closely aligned on the diagram, while the third galaxy in the triplet shows a significantly different $\Delta v/\sigma$. This suggests that these groups might actually consist of one galaxy pair plus an interloper. Among the other two CGs in this cluster, one is located in the intermediate infallers region, while the other lies in the ancient infallers region, indicating that it is already part of the cluster.

Cluster number 119, CGs-2360 show that all group members have similar projected radii but distinct $\Delta v/\sigma$ values, suggesting that these two CGs are likely not genuinely dense systems, but projections, instead. The other CG in this cluster appears to be a system in the process of recent infall. Finally, in cluster number 122, the CGs have galaxies located in the ancient infallers region, indicating that these galaxies are already part of the cluster.
Unfortunately, we cannot perform this type of analysis for the remaining non-isolated CGs because the structures they are embedded in contain fewer members, making it difficult to assume that these major structures are close to dynamical equilibrium.

In summary, within these four clusters, we observe several CGs that are not truly dense structures. This could be due to a selection bias, as both catalogues focus on visually identifying dense projected structures and allow for high velocity differences of up to 1000 km/s. 1000 km/s in the line of sight corresponds to a considerable distance of $\sim 10$ Mpc/h, increases the likelihood of misidentifications (\citealt{2018Diaz}, \citealt{2024Zandivarez}). A more appropriate approach for studying CGs embedded in clusters would involve deliberately searching for them within these systems, starting with pre-identified clusters and applying a substructure detection algorithm that limits velocity differences, in the appendix, we discuss this in more detail (see Section~\ref{app:calsagos}). Additionally, we find systems that are either recently falling into clusters or are intermediate or ancient infallers. Interestingly, the latter may represent remnants of larger groups that fell into the clusters and, as they evolved, lost members, leaving behind only the central part of the group (\citealt{Haggar2022}).

In summary, across the four clusters, we identify several CGs that are compact in projection yet dynamically inconsistent in projected phase–space, more compatible with chance alignments than with bound substructures. This outcome is consistent with a selection bias: both catalogues prioritise visually dense configurations and tolerate large radial–velocity windows (e.g. $|\Delta v|\!\le\!1000~\mathrm{km\,s^{-1}}$), which corresponds to a line–of–sight distance of $\simeq 10\,h^{-1}\,\mathrm{Mpc}$ and increases the risk of misidentifications \citep{2018Diaz,2024Zandivarez}. A more appropriate strategy for CGs embedded in clusters is to start from pre–identified clusters and apply a substructure–finding algorithm that explicitly enforces velocity coherence; we outline such an approach in Appendix~\ref{app:calsagos}. Within the PPSD subset (12 CGs in 4 clusters), we also find systems consistent with recent, intermediate, and ancient infall; the latter may be the surviving cores of larger groups accreted in the past, having lost members during their evolution \citep{Haggar2022}. For non–isolated CGs lacking robust host parameters, we do not extrapolate these PPSD trends.

\section{Discussion}
\label{sec:discussion}

\subsection{Differences in structural parameters between CG and surrounding group galaxies}
\label{sec:discussion_structural}

The variation of $n$ and $R_e$ with wavelength provides insight into how the relative contributions of bulge and disc components change across stellar populations. Blue filters trace the light of young stars and the disc, while redder filters are dominated by older stellar populations in bulges (\citealt{1998Kennicutt}; \citealt{2006Moorthy}). As a result, $n$ typically increases towards longer wavelengths, especially in galaxies with significant bulge–disc structures (\citealt{2014vulcani}). This is precisely what we find for LTGs, which display a stronger wavelength dependence of $n$ than ETGs, the latter being generally well described by single–component models (\citealt{2001Iodice}). Although each galaxy was modelled with a single Sérsic profile, these wavelength-dependent trends still encode information about internal structural and stellar population gradients, as was show by \cite{2023Montaguth}.

Comparing each morphological type between environments (Figure~\ref{fig:forest_cliffs}) reveals at most a weak but consistent environmental imprint. For ETGs, non–isolated CGs show systematically steeper $n(\lambda)$ at intermediate and faint luminosities, while differences in the brightest bin are small and not significant, suggesting that ETG structures respond in broadly similar ways to stellar population gradients in both environments. LTGs likewise show little or no environmental dependence across most bins; only in the brightest bin does $n(\lambda)$ separate markedly between environments, with a large negative effect in non–isolated CGs, although this result is based on a very small CG subsample ($n_{\mathrm{CG}}=4$).

Transition galaxies display a mixed behaviour that likely reflects different evolutionary stages. At intermediate luminosity, $n(\lambda)$ is modestly steeper in non-isolated CGs , consistent with enhanced centralisation within dense environments. In the faintes bin, the trend reverses: surrounding-group galaxies show steeper $n(\lambda)$, suggestive of transformation proceeding outside CG cores or of recently quenched, still diffuse systems in the larger-scale environment. For $R_e(\lambda)$, transition systems show no significant environmental differences. Additionally, we find no galaxies in the brightest bin, in non-isolated CGs and just one galaxy in the surrounding groups sample. One possible explanation for this absence is that the most luminous galaxies may have already undergone morphological transformation and now appear as early-type systems, typically exhibiting Sérsic indices above $n = 2.5$. Another possibility is that, as spiral galaxies lose their arms and become less luminous, they may fall
below detection thresholds, potentially affecting their inclusion in morphological analyses \citep{2006Bedregal}. Overall, the similarity of the wavelength-dependent trends in both environments indicates that the physical mechanisms shaping galaxy structure operate in both non-isolated CGs and their host environments.

Moreover, by examining the relationship between $R_e$ and $n$ in the $r$-band (Figure \ref{fig:re_n_TG}), we identified more clearly the structural differences between transition galaxies in non-isolated CGs and those in surrounding groups. In the intermediate magnitude bin, most transition galaxies in non-isolated CGs have $n \gtrsim 1.5$, suggesting a more advanced stage of morphological transformation, while those in the surrounding groups span a broader range in $n$, with many at $n<1.75$, showing a clear separation from the non-isolated CG sample and indicating a more heterogeneous population.


In the intermediate-luminosity bin, galaxies in non-isolated CGs appear, on average, more morphologically evolved, with higher-n systems than those in the surrounding sample. This is consistent with studies showing that galaxies in Hickson CGs host older stellar populations than comparable field or loose-group galaxies \citep{2004Proctor,2012Plauchu-Frayn,2015Coenda}. Taken together, these results support the view that dense CG environments accelerate both morphological transformation and the build-up of older, quenched stellar populations, in line with our $R_{e}-n$ trends. By contrast, the surrounding sample spans a range of host haloes (from loose groups to cluster outskirts), naturally yielding a broader mix with many systems still transitioning morphologically and therefore showing lower $n$ on average.

In the faint bin, TGs in  non-isolated CGs can plausibly be late-type systems moving towards early-type after gas loss, already redder in colour \citep{2001Verdes, 2023Jones} but still catching up morphologically, which explains the split between low-n and high-n systems. In the surrounding sample, the higher n of faint TGs suggests a population that has already progressed further along this path in groups/clusters. Overall, this supports a picture in which CGs drive faster evolution (cluster-like ages; \citealt{2004Proctor}), yet heterogeneity remains visible among faint TGs because some are mid-transition while others are already structurally transformed.

\subsection{Fraction of quenched galaxies and ETGs as a function of the distance to the centre of the major structure and their local densities}

We identify differences in the fraction of quenched galaxies and ETGs as a function of the stellar masses when comparing galaxies within non-isolated CGs to those in their surrounding groups. In non-isolated CGs, the fraction of quenched galaxies and ETGs is higher than in their surrounding group sample (see right panel in Figure \ref{fig:que_frac}). These differences become particularly significant for galaxies with stellar masses greater than $log(M[M_{\odot}]) = 10.2$. These results are consistent with the findings of \cite{2023Taverna}, who reported that CGs, regardless of the large-scale environment in which they are embedded, systematically show higher fractions of red and early-type galaxies compared to their surroundings.

However, although they compared the fractions of ETGs exclusively within CGs embedded in different environments, no significant variation was observed in the difference between CG and surrounding group galaxy fractions. They associated this trend with processes occurring within the CGs themselves. Nevertheless, these results may also be related to the distance from the centres of the larger structures. In the case of CGs located in more central regions, due to the well-known density–morphology (\citealt{1984Dressler}, \citealt{Baldry2004b}) and density–star formation (\citealt{2003Kauffmann}) relations, it is expected that they show a higher fraction of quenched galaxies and ETGs than the overall galaxy population in their surroundings. For that reason, we divided our sample according to whether the non-isolated CGs or galaxies within the host structures are located inside or outside $R_{200}$ (see Figure \ref{fig:que_frac_dis}).

Interestingly, the higher quenched fractions in non-isolated CGs remain for those systems situated outside of the $R_{200}$ (see the bottom left panel in Figure \ref{fig:que_frac_dis}). This suggests that quenching is enhanced in denser environments, likely due to a combination of environmental mechanisms such as strangulation (\citealt{1972Gunn}), ram-pressure stripping (\citealt{2008Kawata}). Importantly, in low-density environments (bottom-left panel Figure \ref{fig:que_frac_dens}), the quenched fraction of galaxies in surrounding groups converges toward that of the non-isolated CGs. This similarity implies that environmental quenching may act beyond the compact configuration itself and could reflect broader environmental processes operating on scales larger than those of CGs. It also hints at a possible evolutionary connection or shared infall histories among these systems. An example of this is the Blue Infalling Group, which is being transformed by gravitational interactions among its member galaxies, including mergers and the formation of tidal dwarfs, as well as by ram-pressure stripping as it falls into the A1367 cluster (\citealt{2019Fossati}).

Moreover, when comparing the fraction of quenched galaxies with that of ETGs, we observe that for non-isolated CGs located outside $R_{200}$ (see the bottom panels in Figure \ref{fig:que_frac_dis}), the ETG fraction is higher in the two least massive bins. This result supports the scenario in which morphological transformation may precede the quenching of star formation. Similar conclusions were reached by \citet{2024Sampaio}, who, using phase-space analysis, found that galaxies often experience morphological changes before becoming fully quenched. In this context, the enhanced ETG fractions in low-mass bins—where the quenched fractions remain lower—may correspond to galaxies undergoing morphological transformation while still retaining residual star formation. Interestingly, this effect is more clearly observed outside $R_{200}$, where the environment is less dense and the transformation process appears to be more gradual. In contrast, within $R_{200}$, the denser environment and the internal dynamics of CGs may have led to an earlier onset of quenching, providing galaxies with enough time to fully complete the process.

The analysis of local density further highlights the complexity of environmental influence (see Figure \ref{fig:que_frac_dens}). In high-density environments, CGs exhibit higher quenched fractions compared to isolated CGs, but the difference with surrounding group galaxies is reduced. In contrast, at low local densities, the quenched fractions in surrounding groups approach those of CGs, suggesting that environmental quenching operates effectively even outside the CG configuration. This may reflect shared infall histories, or the influence of the larger-scale structure, consistent with the concept of galactic conformity (\citealt{2015Kauffmann}; \citealt{2023Ayromlou}).


\section{Summary and Conclusions}
\label{sec:conclusions}

In this study, we analysed data from 122 compact groups (CGs) embedded within 102 major structures, classifying these CGs as non-isolated. Galaxies within these major structures that are not identified as CG members are designated as surrounding group galaxies. The aim of this work is to investigate the structural and dynamical properties of galaxies within non-isolated CGs and their surrounding environments. In the analysis, we use multi-wavelength observations from the S-PLUS survey (\citealt{mendes2019southern}), which includes 12 optical filters. Using this data set and the \textsc{morphoPLUS} code, we estimate the structural parameters of each galaxy. This information was further complemented with the GSWL catalogue (\citealt{2018GSWLC}) to obtain the star formation rates (SFRs) and the stellar masses. The galaxy populations in each sample were classified into early-type (ETGs), late-type (LTGs), and transition galaxies based on their Sérsic indices ($n$) and colours (\citealt{vika2015megamorph}). In addition, to analysing morphology, we divided the galaxy sample into three bins based on absolute magnitudes in the $r$-band: the brightest galaxies with magnitudes $-23.7 < M_r \leq -22.3$, intermediate magnitudes $-22.3 < M_r \leq -20.8$, and fainter galaxies with magnitudes $-20.8 < M_r \leq -18$. We also distinguish galaxies in the surrounding groups and in non-isolated CGs between high-density (log$(\Sigma_5)>$-2.1) and low-density (log$(\Sigma_5)\leq$-2.1) environments.
\begin{itemize}

    \item In general, $n$ increases with wavelength for all types, consistent with a growing contribution from centrally concentrated, older stellar populations in redder bands. The behaviour of $R_e$, however, is more complex. For ETGs, $R_e$ increases with wavelength among the brightest systems, while fainter ones—especially in non-isolated CGs, show a decrease, suggesting more compact red stellar components. LTGs exhibit decreasing $R_e$ with wavelength, particularly in surrounding groups, likely reflecting fading star-forming disks in redder bands. We find no significant differences between environments; the similarity of the wavelength-dependent trends suggests that the mechanisms shaping galaxy structure operate in both compact groups and larger group setting.
    
    \item The $R_e$–$n$ analysis reveals significant structural differences between transition galaxies in non-isolated CGs and those in surrounding groups. In the intermediate magnitude bin, galaxies in CGs are more compact and exhibit higher Sérsic indices, suggesting more advanced morphological evolution. In the faint bin, the bimodality seen in non-isolated CGs, absent in surrounding groups, indicates ongoing transformation, while the surrounding galaxies appear more homogeneous and consistent with post-transformation early types. Interestingly, the bimodality reported in \cite{2024Montaguth} for non-isolated CGs is primarily driven by faint galaxies, whereas the brightest ones contribute mainly to the high-$n$ peak of the distribution. These trends, confirmed by the Mahalanobis distance test, highlight the role of environment in shaping galaxy structure, especially in low-luminosity systems.

    \item Non-isolated CG exhibit a higher fraction of quenched galaxies and a greater proportion of ETGs compared to their surroundings, particularly for galaxies with stellar masses $\log(M/M_\odot) > 10.2$. However, these differences become less pronounced when the sample is divided based on whether galaxies are located inside or outside the $R_{200}$ of the major structures, or when classified according to their local densities as low or high. This suggests that both quenching and morphological transformation are strongly linked to the distance from the center of major structures, and to the local environment. Although the compact nature of the non-isolated CGs appears to enhance these processes, surrounding groups exhibit similar evolutionary trends under certain conditions. This emphasizes that the evolution of non-isolated CGs is a product of both their compact nature and the major structures in which they are embedded, reforging the concept of galactic conformity.
    
    \item Non-isolated CGs exhibit a higher fraction of quenched galaxies and a greater proportion of ETGs compared to their surroundings, especially for galaxies with $\log(M/M_\odot) > 10.2$. These differences are less pronounced when splitting the sample by location relative to $R_{200}$ or by local density, suggesting that both quenching and morphological transformation are influenced by environment and distance to the centre of the host structure. Notably, outside $R_{200}$, the ETG fraction exceeds the quenched fraction in the two lowest mass bins, supporting the scenario in which morphological transformation may precede star formation quenching under specific conditions.

    \item  Our dynamical analysis of CGs embedded in clusters indicates that several of these systems are not truly physically dense, but rather chance alignments in projection. This interpretation is supported by the significant differences in their line-of-sight velocities relative to the cluster centre, despite their apparent spatial proximity. Additionally, we identify CGs across various stages of infall, from recent arrivals to long-embedded systems. Notably, ancient infallers may correspond to the remnants of more massive groups that fell into the cluster and gradually lost members due to environmental processes, leaving only the dense core. These findings highlight the need for more robust identification techniques for CGs within clusters, particularly those incorporating substructure detection methods and tighter velocity constraints to reduce contamination by projection effects.


\end{itemize}

In summary, our findings confirm previous results from \cite{2024Montaguth}, showing that the evolution of galaxies in non-isolated CGs is affected by their location within larger structures. However, this study goes beyond by directly comparing galaxies in non-isolated CGs to those in their surrounding groups. We find that the structural and star-forming properties of galaxies in non-isolated CGs differ significantly from those of their neighbours in the host structures.
These differences suggest that galaxies in non-isolated CGs are evolving through distinct environmental mechanisms, likely driven by the compactness and local interactions within CGs themselves. While our current analysis provides valuable insights, it is limited by the number of systems. As a natural next step, we plan to identify and characterise new CGs in the southern sky using the S-PLUS survey. Expanding the sample will be essential to test and refine the trends presented here.

\begin{acknowledgements}

We are grateful to the anonymous referee for their constructive comments, which have significantly improved this paper. G.M. gratefully acknowledges the Fundação de Amparo à Pesquisa do Estado de São Paulo (FAPESP) for the support grant 2024/10923-3. CL-D acknowledges financial support from the ESO Comite Mixto 2022. ST-F acknowledges the financial support of ULS/DIDULS through a regular project number PR2453858. AM acknowledges support from the ANID FONDECYT Regular grant 1251882,  from the ANID BASAL project FB210003, and funding from the HORIZON-MSCA-2021-SE-01 Research and Innovation Programme under the Marie Sklodowska-Curie grant agreement number 101086388. 

DP gratefully acknowledge financial support from ANID - MILENIO – NCN2024\_112. DP acknowledges financial support from ANID through FONDECYT Postdoctorado Project 3230379 and  FONDECYT Regular grants no. 1241426. P.K.H. gratefully acknowledges the Fundação de Amparo à Pesquisa do Estado de São Paulo (FAPESP) for the support grant 2023/14272-4. SP is supported by the international Gemini Observatory, a program of NSF NOIRLab, which is managed by the Association of Universities for Research in Astronomy (AURA) under a cooperative agreement with the U.S. National Science Foundation, on behalf of the Gemini partnership of Argentina, Brazil, Canada, Chile, the Republic of Korea, and the United States of America.
R.D. gratefully acknowledges support by the ANID BASAL project FB210003. AAC acknowledges financial support from the Severo Ochoa grant CEX2021- 001131-S funded by MCIN/AEI/10.13039/501100011033 and from the project PID2023-153123NB-I00, funded by MCIN/AEI. R.F.H acknowledge financial support from Consejo Nacional de Investigaciones Cientificas y Técnicas (CONICET) (PIP 1504), Agencia I+D+i (PICT 2019–03299) and Universidad Nacional de La Plata (Argentina).

The S-PLUS project, including the T80-South robotic telescope and the S-PLUS scientific survey, was founded as a partnership between the Fundação de Amparo à Pesquisa do Estado de São Paulo (FAPESP), the Observatório Nacional (ON), the Federal University of Sergipe (UFS), and the Federal University of Santa Catarina (UFSC), with important financial and practical contributions from other collaborating institutes in Brazil, Chile (Universidad de La Serena), and Spain (Centro de Estudios de Física del Cosmos de Aragón, CEFCA). We further acknowledge financial support from the São Paulo Research Foundation (FAPESP), the Brazilian National Research Council (CNPq), the Coordination for the Improvement of Higher Education Personnel (CAPES), the Carlos Chagas Filho Rio de Janeiro State Research Foundation (FAPERJ), and the Brazilian Innovation Agency (FINEP).

The authors are grateful for the contributions of CTIO staff in helping in the construction, commissioning and maintenance of the T80-South telescope and camera. We are also indebted to Rene Laporte and INPE, as well as Keith Taylor, for their important contributions to the project. We also thank CEFCA staff for their help with T80-South. Specifically, we thank Antonio Marín-Franch for his invaluable contributions in the early phases of the project, David Cristóbal-Hornillos and his team for their help with the ective radius (installation of the data reduction package jype version 0.9.9, César Íñiguez for providing 2D measurements of the filter transmissions, and all other staff members for their support.
\end{acknowledgements}

%
%

\bibliography{sample701}{}
\bibliographystyle{aasjournalv7}
\begin{appendix}

\section{MorphoPLUS: the code explanation}
\label{app:morphoplus}

\begin{figure*}
    \centering
    \includegraphics[width=0.9\textwidth]{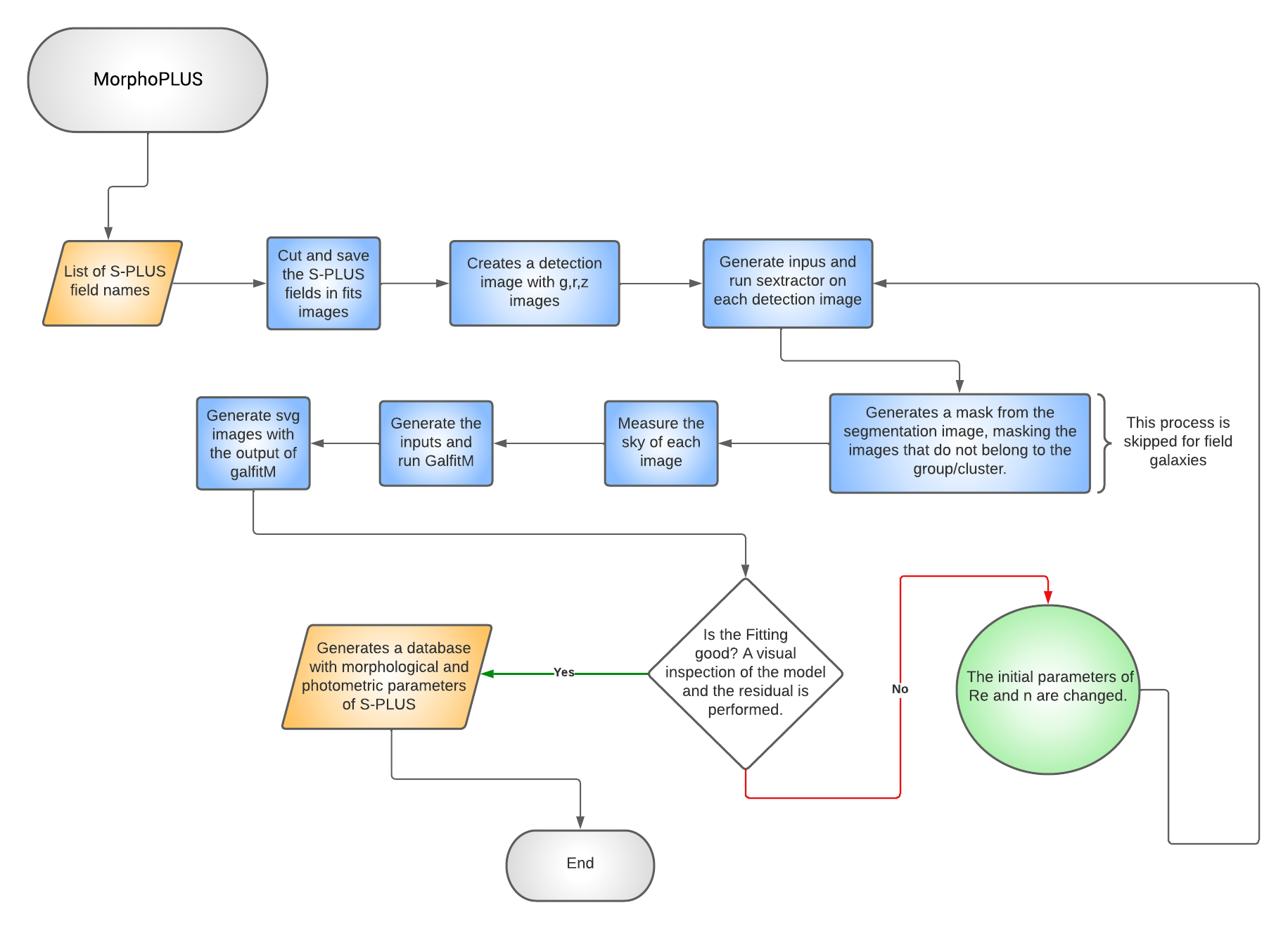}
    \caption{A flowchart describing \textsc{MorphoPLUS} operation. Required inputs are the catalogue with a list of galaxy coordinates, and magnitudes.}
    \label{fig:Dia_flu}
\end{figure*}

We developed a code that we named `\textsc{MorphoPLUS}', which, as mentioned in the methodology section, is based on \textsc{Bash} and \textsc{Python}, utilises \textsc{SExtractor} tools, and is built upon the \textsc{MegaMorph} framework. \textsc{MegaMorph} facilitates the extraction of morphometric parameters for each galaxy and performs two-dimensional fitting across multiple wavelengths using the \textsc{GALFITM} algorithm, a modified version of \textsc{GALFIT} 3.02 \citep{2002Peng, 2010Peng}.

\textsc{GALFITM} models the surface brightness of a galaxy using a two-dimensional analytical function, which can be Sérsic, Nuker, de Vaucouleurs, or exponential (\citealt{Sersic}, \citealt{1995lauer}, \citealt{1948Vaucouleurs}, \citealt{1970Freeman}). To extend these functions across multiple wavelengths, free parameters are replaced by wavelength-dependent functions represented by a set of Chebyshev polynomials. Additionally, \textsc{GALFITM} allows for multi-component fitting, enabling users to include information about the sky background and decompose the galaxy into its bulge, disk, and bar components. The best fit is determined using the Levenberg-Marquardt (LM) algorithm, which minimises $\chi^2$.

Essentially, \textsc{MorphoPLUS} is automated to take a catalogue as input, containing a list of objects for fitting using S-PLUS data. This catalogue includes the RA and Dec coordinates, magnitudes in all 12 filters,  half-light radius ($R_{50}$), the radius containing 90\% of the light ($R_{90}$), axis ratio, and the X and Y pixel positions of the galaxies in the S-PLUS images, information that is already provided within the S-PLUS catalogues.

In Figure \ref{fig:Dia_flu}, we present the flowchart of the code. The process begins by identifying the corresponding S-PLUS fields for the groups under analysis. It then locates the galaxies in these fields and divides each field into sub-images in the 12 filters. This division is necessary because the S-PLUS images are very large, and splitting them into smaller sub-images allows for more efficient processing and fitting with \textsc{GALFITM}. It is worth noting that, due to the automated process, some galaxies may extend to the edges of these subfields. However, the code is designed to identify and exclude these edge galaxies from the fitting process. Subsequently, the code generates images for these edge galaxies that are four times larger than the original sub-images, allowing for proper fitting in a later step.

In addition to creating and saving the images, the code also generates the point spread functions (PSFs) for each filter. These PSFs are modelled using a Moffat function (\citealt{1969Moffat}), defined as:

\begin{equation}
PSF(r) = \frac{\beta - 1}{\pi \alpha^2} \left[ 1 + \left( \frac{r}{\alpha} \right)^2 \right]^{-\beta}
\end{equation}

with full width at half maximum (FWHM) \( \text{FWHM} = 2\alpha \sqrt{2^{1/\beta} - 1} \). Both the FWHM and the beta parameter (\(\beta\)) are available in the headers of each field for every filter in the S-PLUS images.

Once the images of each subfield and individual galaxy are prepared, the code proceeds to generate the detection images, which are created by combining the $g$, $r$, and $z$ filters. These images are then used to produce segmentation images with the \textsc{SExtractor} software. \textsc{SExtractor} masks all detected sources, including the galaxies in the input catalogue.

Next, the code uses these images to generate a new segmentation image, in which it unmasks only the galaxies from the provided catalogue. In this process, the code assigns a numerical value of one to masked pixels and zero to unmasked pixels. This step is crucial, as \textsc{GALFITM} requires the input mask to contain zeros for the regions to be fitted.

Afterwards, the code measures the background sky value for each image by calculating the median of pixel values at a $3\sigma$ level, effectively excluding potential sources. It is important to note that S-PLUS images already have the background sky-subtracted.  

With these estimated values, the code proceeds to generate input parameters for each subfield. It requires an initial guess for the parameters to be modelled, which are taken from the S-PLUS catalogues. These include: $R_{50}$, axis ratio, apparent magnitudes in the twelve filters, and zero points of the images. The zero points are available in additional tables within the S-PLUS catalogues\footnote{The zero points can be found in the auxiliary tables section: \url{https://splus.cloud}}.  

For the initial Sérsic index ($n$) estimate, the code applies an empirical relation between concentration ($C$), defined as the ratio $R_{90}/R_{50}$, and $n$ \citep{2011Andrae}, given by:  

\begin{equation}
    C = 2.770n^{0.466}
\end{equation}

Then the code runs \textsc{GALFITM}, which, in this case, fits all galaxies with a single component model using a Sérsic profile, as shown in equation \ref{eq:sersic}:

\begin{equation}
    \centering
    I=I_e \exp{\left[-b_n \left(\left(\frac{R}{R_e}\right)^{1/n}-1\right)\right]}
    \label{eq:sersic}
\end{equation}

Here, the intensity at the effective radius ($R_e$), denoted as $I_e$, is a key parameter. It represents the intensity at the radius containing half of the total light. The parameter $b_n$ is a function of the Sérsic index ($n$) and has a relationship defined as $\Gamma(2n) = 2\gamma(2n, b_n)$, where $\Gamma$ represents the Gamma function, and $\gamma$ denotes the lower incomplete Gamma function (\citealt{1991Ciotti}). 

Then, the code generates two tables: one containing fits with $\chi^2 < 2$, and another with fits where $\chi^2 > 2$. Additionally, the code creates SVG images displaying each subfield in the twelve filters, along with their corresponding models and residuals.

For galaxies with $\chi^2 > 2$, we refine the models to account for potential errors, such as contamination from a nearby star or interference from a bright neighbouring source affecting the mask image. Therefore, we adjust the image masks individually for these cases and rerun \textsc{GALFITM}.

We also perform a visual inspection of the fits to ensure quality control. Specifically, we verify galaxies with $\chi^2 < 2$ to confirm that the models are appropriate. Additionally, we rerun \textsc{GALFITM} for galaxies where the visual inspection indicated an unsatisfactory fit.

\begin{figure*}

    \includegraphics[scale=0.3]{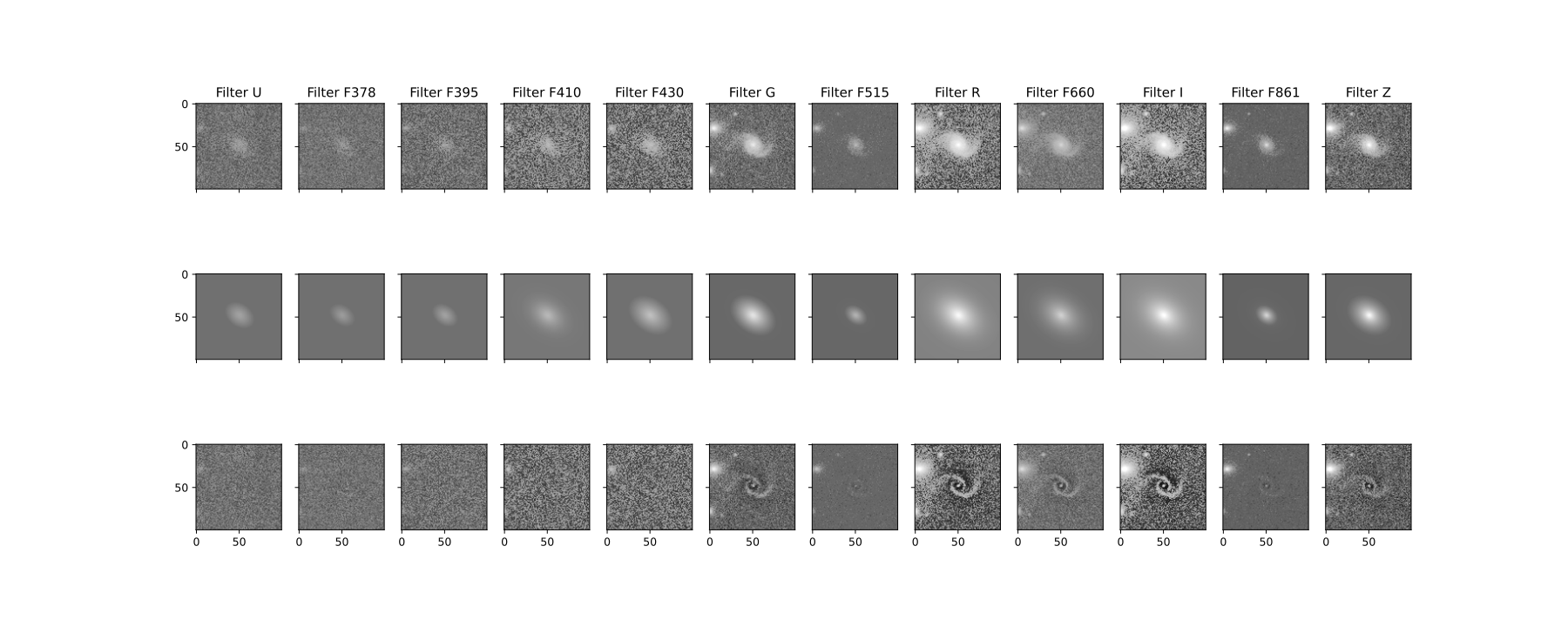}
    \caption{Upper panels: S-PLUS images for one of the CGs analysed in each filter, as labeled at the top of each panel, and the borders are in pixels. Middle panels: the model generated using \textsc{MorphoSPLUS}. bottom panels: the residual, the model subtracted from the observed image.}
   \label{fig:215}
\end{figure*}

In summary, the \textsc{MorphoPLUS} code is designed to automate the modelling of galaxy surface brightess using a single-component Sérsic profile for S-PLUS data. Users only need to provide a table containing the catalogue of galaxies of interest. Previously, \textsc{MorphoPLUS} focused on CGs, using images centred on CG coordinates \citep{2023Montaguth}. However, for larger systems, such as loose groups or clusters, we modified the pipeline to divide the S-PLUS fields into smaller sub-images, improving computational efficiency. In Figure \ref{fig:215}, we present an example of an output for one of the galaxies analysed in this work. The top panels display the images in each S-PLUS filter, the middle panels show the light profile model, and the bottom panels present the residual image, namely, the observed image with the model subtracted.

\section{Galaxy classification}
\label{app:morph_cuts}

Galaxies populate a well-known colour–structure plane: star-forming discs are typically blue with low Sérsic indices, while quiescent spheroids are red with high 
$n$ \citep{morgan1957spectral,2008Ball,2014vulcani,vika2015megamorph}. We characterise each galaxy by its rest-frame $(u-r)_0$(Galactic-extinction and k-corrected) and the single-component Sérsic index $n_r$ in the r band. Rather than adopting fixed cuts, we derive boundaries from the data and then compare them with parameters $(u-r)=2.3$, and $n_r=2.5$ estimated by \cite{vika2015megamorph}.

\begin{figure}
    \centering
    \includegraphics[scale=0.5]{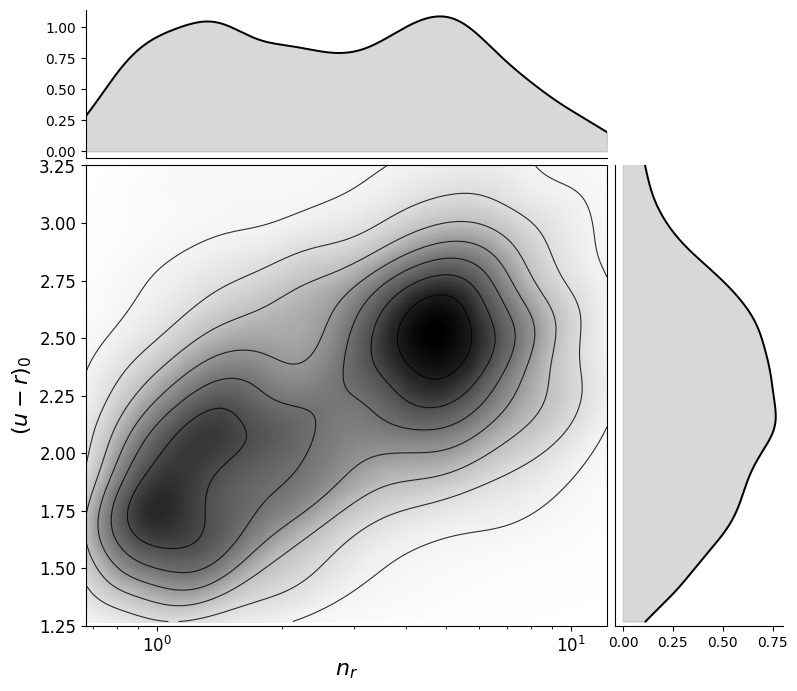}
    \caption{ Two–dimensional kernel–density estimate (KDE) of the $(u-r)_0$- log⁡($n_r$) plane for the group sample augmented with the homogeneous field control set (processed with the same pipeline). Contours show KDE isodensity levels (50, 80, and 95 per cent) computed with a Gaussian kernel, revealing the characteristic bimodality into blue/low–n and red/high–n concentrations. The marginal distributions are displayed as normalised histograms.}
   \label{fig:u-r-total}
\end{figure}

\begin{figure}
    \centering
    \includegraphics[scale=0.5]{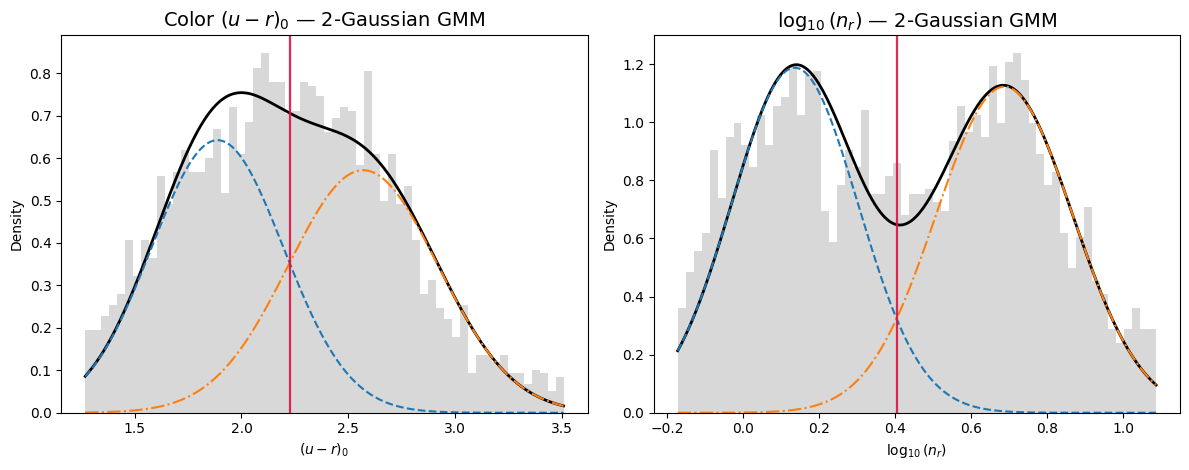}
    \caption{Two–component Gaussian Mixture fits to the one–dimensional distributions of $(u-r)_0$ colour (left) and  log⁡($n_r$) (right). Grey bars show normalised histograms. The total mixture PDF is plotted as a solid black curve, and the two Gaussian components as dashed and dot–dashed curves. Red vertical lines mark the equal–contribution points (intersections of the weighted components), which define our data–driven boundaries: $(u-r)_0=2.22$ and $n_r=2.54$.}
   \label{fig:ggm}
\end{figure}

We complement the group sample with a homogeneous field control set ($\sim 2200$ galaxies) analysed with the same pipeline \citep{2023Montaguth}, in order to increase the statistical power and to obtain better sampling of the different morphological types. In Figure \ref{fig:u-r-total}, we show the galaxy density distribution in the 
$(u-r)_0$-$log⁡(n_r)$ plane, together with the one-dimensional distributions of each parameter in the marginal panels. This figure reveals two concentrations—blue/low-n and red/high-n—as expected for discs ($n\sim1$) and ellipticals ($n\sim4$). We model the 1-D distributions of $(u-r)_0$ and log⁡($n_r$) using two-component Gaussian Mixture Models (GMMs) fitted to the unbinned data, as shown in Figure \ref{fig:ggm}, we present the distributions of both parameters together with the two Gaussian components that provide the best fit: for the colour in the left panel and for $n_r$ in the right panel. For each parameter, the objective threshold is the abscissa where the two component PDFs are equal. This yields $(u-r)_0=2.22$ and 
$n_r=2.54$, which are represented by the red lines in each panel in Figure \ref{fig:ggm}. The colour boundary is slightly bluer than Vika’s 2.3, consistent with our explicit extinction and k-corrections. For practical use, we adopt rounded limits 
$(u-r)_0=2.2$ and $n_r=2.5$.

\section{Substructures inside of the clusters}
\label{app:calsagos}

To identify substructures in the four clusters analysed in this work, we employed the \textsc{CALSAGOS} package (\citealt{2022Olave-Rojas}), which combines clustering algorithms to detect spatial and dynamical overdensities within clusters. We identified the substructures using density-based spatial clustering of applications with noise which was applied to the projected distribution of galaxies within the redshift range of the cluster to detect spatial overdensities. Following the standard configuration, we required a minimum of three galaxies to define a substructure and used the distribution of distances to the third-nearest neighbour to estimate the search radius (eps) parameter. Due to the lower number densities in the outer regions, the search was adapted by estimating eps separately inside and outside $R_{200}$. Specifically, for galaxies within $R_{200}$, distances were computed using only galaxies located inside that radius, while for the outskirts, the full sample was used. This strategy allowed for a more robust identification of substructures across different density regimes.

\begin{figure}
    \centering
    \includegraphics[scale=0.35]{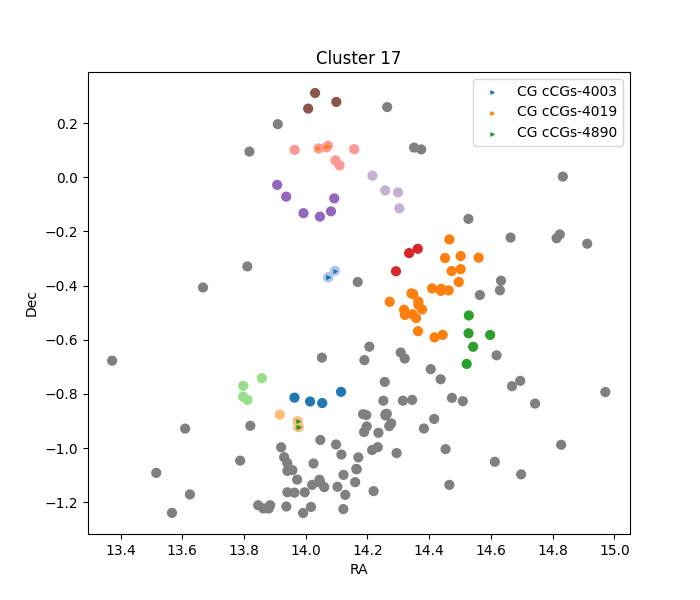}
    \includegraphics[scale=0.35]{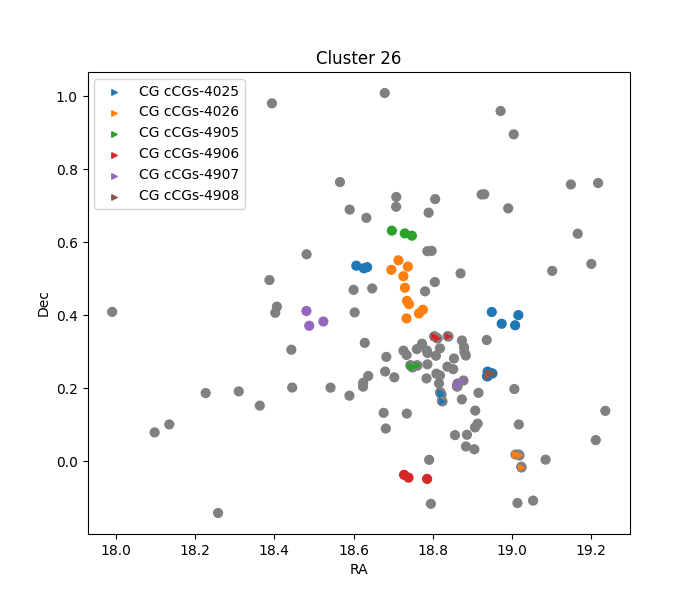}
    \includegraphics[scale=0.35]{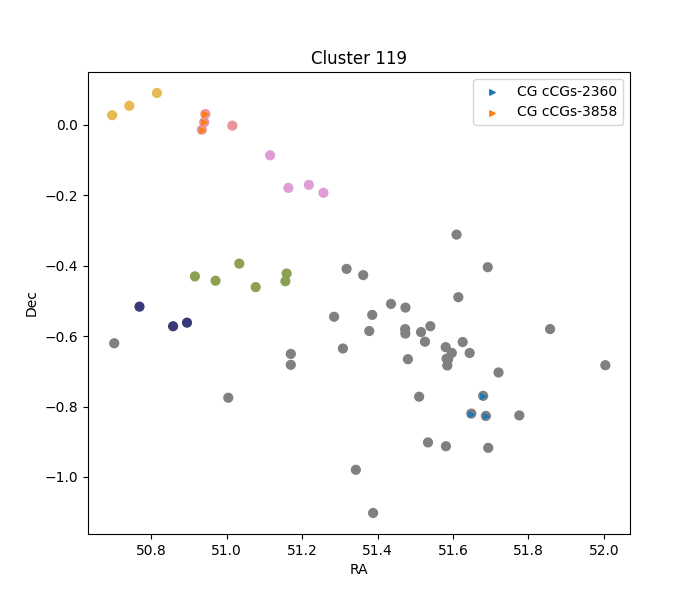}
    \includegraphics[scale=0.35]{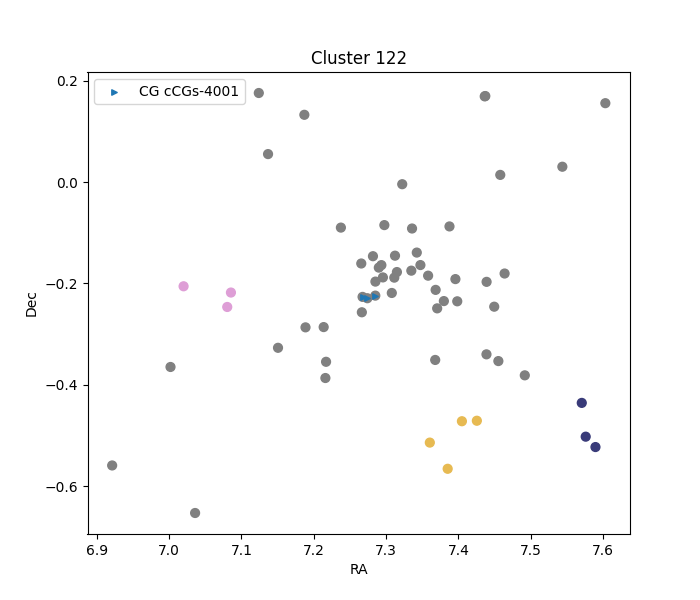}
    \caption{Spatial distribution of galaxies in the four analysed clusters. Each point represents a galaxy: coloured points correspond to galaxies identified as part of a substructure by CALSAGOS, while grey points represent galaxies not assigned to any substructure. Triangles indicate galaxies that belong to non-isolated CGs.}
   \label{fig:calsagos}
\end{figure}

In Figure \ref{fig:calsagos}, the points represent galaxies in the clusters. Galaxies identified as part of a substructure are shown in different colours, while those not associated with any substructure are shown in grey. Triangles overlaid on the points indicate which galaxies were identified as members of non-isolated CGs. When comparing these figures with PPSDs, we find that, interestingly, in the case of cluster 17, the non-isolated CGs coincide with substructures identified using \textsc{CALSAGOS}, and in the PPSD, their members are confined within the cluster's velocity dispersion range, suggesting that they are indeed genuine substructures. However, using \textsc{CALSAGOS}, we find that two of the three CGs in this cluster are part of even larger substructures. In cluster 26, only one CG overlaps with a detected substructure, suggesting that this system may represent a genuine substructure. In cluster 119, CG-4908 coincides with a substructure identified by \textsc{CALSAGOS}; however, an inspection of the projected phase-space diagram reveals that one of its galaxies exhibits a velocity significantly different from the other two, raising doubts about the physical reality of the system. In cluster 122, the CG does not correspond to any overdensity, further indicating that it is likely a chance alignment rather than a bound group.

These results highlight the importance of employing a refined identification strategy when studying CGs within clusters. Instead of relying exclusively on Hickson’s criteria, we propose first identifying substructures using objective clustering algorithms, and subsequently applying the Hickson criteria to determine which of these substructures may correspond to embedded CGs. This two-step approach helps to avoid biased or misleading classifications that can arise from projection effects or chance alignments.
\end{appendix}

%
%



\end{document}